\input epsf

\def\figin{\epsfcheck\figin}\def\figins{\epsfcheck\figins}
\def\epsfcheck{\ifx\epsfbox\UnDeFiNeD
\message{(NO epsf.tex, FIGURES WILL BE IGNORED)}
\gdef\figin##1{\vskip2in}\gdef\figins##1{\hskip.5in}
\else\message{(FIGURES WILL BE INCLUDED)}%
\gdef\figin##1{##1}\gdef\figins##1{##1}\fi}
\def\DefWarn#1{}
\def\figinsert{\goodbreak\topinsert}
\def\ifig#1#2#3#4{\DefWarn#1\xdef#1{fig.~\the\figno}
\writedef{#1\leftbracket fig.\noexpand~\the\figno}%
\figinsert\figin{\centerline{\epsfxsize=#3mm \epsfbox{#2}}}
\bigskip\medskip\centerline{\vbox{\baselineskip12pt
\advance\hsize by -1truein\noindent\footnotefont{\sl Fig.~\the\figno:}\sl\ #4}}
\bigskip\endinsert\noindent\global\advance\figno by1}

\newcount\figno
 \figno=1
 \def\fig#1#2#3{
 \par\begingroup\parindent=0pt\leftskip=1cm\rightskip=1cm\parindent=0pt
 \baselineskip=11pt
 \global\advance\figno by 1
 \midinsert
 \epsfxsize=#3
 \centerline{\epsfbox{#2}}
 \vskip 12pt
 {\bf Fig.\ \the\figno: } #1\par
 \endinsert\endgroup\par
 }
 \def\figlabel#1{\xdef#1{\the\figno}}
 \def\encadremath#1{\vbox{\hrule\hbox{\vrule\kern8pt\vbox{\kern8pt
 \hbox{$\displaystyle #1$}\kern8pt}
 \kern8pt\vrule}\hrule}}

\def\boxit#1{\vbox{\hrule\hbox{\vrule\kern8pt
\vbox{\hbox{\kern8pt}\hbox{\vbox{#1}}\hbox{\kern8pt}}
\kern8pt\vrule}\hrule}}
\def\mathboxit#1{\vbox{\hrule\hbox{\vrule\kern8pt\vbox{\kern8pt
\hbox{$\displaystyle #1$}\kern8pt}\kern8pt\vrule}\hrule}}

\def\half{{1\over 2}}
\def\wtl{{\widetilde\lambda}}
\def\lt{{\widetilde\lambda}}
\def\a{{\alpha}}

\def\b{{\beta}}
\def\d{{\delta}}
\def\g{{\gamma}}
\def\G{{\Gamma}}
\def\e{{\epsilon}}
\def\z{{\zeta}}
\def\ve{{\varepsilon}}

\def\m{{\mu}}

\def\u{{\Upsilon}}
\def\l{{\lambda}}
\def\s{{\sigma}}
\def\t{{\sigma}}
\def\vt{{\theta}}


\def\CA{{\cal A}}
\def\CB{{\cal B}}
\def\CC{{\cal C}}
\def\CD{{\cal D}}

\def\CF{{\cal F}}

\def\CH{{\cal H}}

\def\CL{{\cal L}}
\def\CM{{\cal M}}
\def\CN{{\cal N}}
\def\CO{{\cal O}}

\def\CS{{\cal S}}
\def\CT{{\cal T}}

\def\CX{{\cal X}}

\def\CZ{{\cal Z}}


\def\bC{{\bf C}}

\def\be{{\bf e}}

\def\bg{{\bf g}}
\def\bH{{\bf H}}

\def\bP{{\bf P}}

\def\bR{{\bf R}}
\def\bS{{\bf S}}
\def\bT{{\bf T}}

\def\bV{{\bf V}}

\def\bZ{{\bf Z}}

\def\p{\partial}
\def\pb{\bar{\partial}}


%
%
%
%
\def\unredoffs{} \def\redoffs{\voffset=-.40truein\hoffset=-.40truein}
\def\speclscape{}
%
%
%
%
\newbox\leftpage \newdimen\fullhsize \newdimen\hstitle \newdimen\hsbody
\tolerance=1000\hfuzz=2pt
\catcode`\@=11 
\def\bigans{b }
\def\answ{b }
\ifx\answ\bigans\message{(This will come out unreduced.}
\magnification=1200\unredoffs\baselineskip=16pt plus 2pt minus 1pt
\hsbody=\hsize \hstitle=\hsize 
\else\message{(This will be reduced.} \let\l@r=L
\magnification=1200\baselineskip=16pt plus 2pt minus 1pt
\vsize=7truein \redoffs
\hstitle=8truein\hsbody=4.75truein\fullhsize=10truein\hsize=\hsbody
\output={\ifnum\pageno=0 
    \shipout\vbox{\speclscape{\hsize\fullhsize\makeheadline}
      \hbox to \fullhsize{\hfill\pagebody\hfill}}\advancepageno
    \else
    \almostshipout{\leftline{\vbox{\pagebody\makefootline}}}\advancepageno
    \fi}
\def\almostshipout#1{\if L\l@r \count1=1 \message{[\the\count0.\the\count1]}
        \global\setbox\leftpage=#1 \global\let\l@r=R
   \else \count1=2
    \shipout\vbox{\speclscape{\hsize\fullhsize\makeheadline}
        \hbox to\fullhsize{\box\leftpage\hfil#1}}  \global\let\l@r=L\fi}
\fi
%
\newcount\yearltd\yearltd=\year\advance\yearltd by -1900

\def\Title#1#2{\nopagenumbers\abstractfont\hsize=\hstitle\rightline{#1}%
\vskip 1in\centerline{\titlefont #2}\abstractfont\vskip .5in\pageno=0}
\def\Date#1{\vfill\leftline{#1}\tenpoint\supereject\global\hsize=\hsbody%
\footline={\hss\tenrm\folio\hss}}
%

\def\draftmode{\message{ DRAFTMODE }\def\draftdate{{\rm preliminary draft:
\number\month/\number\day/\number\yearltd\ \ \hourmin}}%
\headline={\hfil\draftdate}\writelabels\baselineskip=20pt plus 2pt
minus 2pt
   {\count255=\time\divide\count255 by 60 \xdef\hourmin{\number\count255}
    \multiply\count255 by-60\advance\count255 by\time
    \xdef\hourmin{\hourmin:\ifnum\count255<10 0\fi\the\count255}}}
\def\nolabels{\def\wrlabeL##1{}\def\eqlabeL##1{}\def\reflabeL##1{}}
\def\writelabels{\def\wrlabeL##1{\leavevmode\vadjust{\rlap{\smash%
{\line{{\escapechar=` \hfill\rlap{\sevenrm\hskip.03in\string##1}}}}}}}%
\def\eqlabeL##1{{\escapechar-1\rlap{\sevenrm\hskip.05in\string##1}}}%
\def\reflabeL##1{\noexpand\llap{\noexpand\sevenrm\string\string\string##1}}}
\nolabels
%
\global\newcount\secno \global\secno=0 \global\newcount\meqno
\global\meqno=1
\def\newsec#1{\global\advance\secno by1\message{(\the\secno. #1)}
\global\subsecno=0\eqnres@t\noindent{\bf\the\secno. #1}
\writetoca{{\secsym} {#1}}\par\nobreak\medskip\nobreak}
\def\eqnres@t{\xdef\secsym{\the\secno.}\global\meqno=1\bigbreak\bigskip}
\def\sequentialequations{\def\eqnres@t{\bigbreak}}\xdef\secsym{}
\global\newcount\subsecno \global\subsecno=0
\def\subsec#1{\global\advance\subsecno by1\message{(\secsym\the\subsecno. #1)}
\ifnum\lastpenalty>9000\else\bigbreak\fi
\noindent{\it\secsym\the\subsecno. #1}\writetoca{\string\quad
{\secsym\the\subsecno.} {#1}}\par\nobreak\medskip\nobreak}
\def\appendix#1#2{\global\meqno=1\global\subsecno=0\xdef\secsym{\hbox{#1.}}
\bigbreak\bigskip\noindent{\bf Appendix #1. #2}\message{(#1. #2)}
\writetoca{Appendix {#1.} {#2}}\par\nobreak\medskip\nobreak}
%
%
\def\eqnn#1{\xdef #1{(\secsym\the\meqno)}\writedef{#1\leftbracket#1}%
\global\advance\meqno by1\wrlabeL#1}
\def\eqna#1{\xdef #1##1{\hbox{$(\secsym\the\meqno##1)$}}
\writedef{#1\numbersign1\leftbracket#1{\numbersign1}}%
\global\advance\meqno by1\wrlabeL{#1$\{\}$}}
\def\eqn#1#2{\xdef #1{(\secsym\the\meqno)}\writedef{#1\leftbracket#1}%
\global\advance\meqno by1$$#2\eqno#1\eqlabeL#1$$}
%
\newskip\footskip\footskip14pt plus 1pt minus 1pt 
\def\footnotefont{\ninepoint}\def\f@t#1{\footnotefont #1\@foot}
\def\f@@t{\baselineskip\footskip\bgroup\footnotefont\aftergroup\@foot\let\next}
\setbox\strutbox=\hbox{\vrule height9.5pt depth4.5pt width0pt}
\global\newcount\ftno \global\ftno=0
\def\foot{\global\advance\ftno by1\footnote{$^{\the\ftno}$}}
%
\newwrite\ftfile
\def\footend{\def\foot{\global\advance\ftno by1\chardef\wfile=\ftfile
$^{\the\ftno}$\ifnum\ftno=1\immediate\openout\ftfile=foots.tmp\fi%
\immediate\write\ftfile{\noexpand\smallskip%
\noexpand\item{f\the\ftno:\ }\pctsign}\findarg}%
\def\footatend{\vfill\eject\immediate\closeout\ftfile{\parindent=20pt
\centerline{\bf Footnotes}\nobreak\bigskip\input foots.tmp }}}
\def\footatend{}
%
%
\global\newcount\refno \global\refno=1
\newwrite\rfile
\def\ref{[\the\refno]\nref}
\def\nref#1{\xdef#1{[\the\refno]}\writedef{#1\leftbracket#1}%
\ifnum\refno=1\immediate\openout\rfile=refs.tmp\fi
\global\advance\refno by1\chardef\wfile=\rfile\immediate
\write\rfile{\noexpand\item{#1\
}\reflabeL{#1\hskip.31in}\pctsign}\findarg}
\def\findarg#1#{\begingroup\obeylines\newlinechar=`\^^M\pass@rg}
{\obeylines\gdef\pass@rg#1{\writ@line\relax #1^^M\hbox{}^^M}%
\gdef\writ@line#1^^M{\expandafter\toks0\expandafter{\striprel@x #1}%
\edef\next{\the\toks0}\ifx\next\em@rk\let\next=\endgroup\else\ifx\next\empty%
\else\immediate\write\wfile{\the\toks0}\fi\let\next=\writ@line\fi\next\relax}}
\def\striprel@x#1{} \def\em@rk{\hbox{}}
\def\lref{\begingroup\obeylines\lr@f}
\def\lr@f#1#2{\gdef#1{\ref#1{#2}}\endgroup\unskip}
\def\semi{;\hfil\break}
\def\addref#1{\immediate\write\rfile{\noexpand\item{}#1}} 
\def\startrefs#1{\immediate\openout\rfile=refs.tmp\refno=#1}
\def\xref{\expandafter\xr@f}\def\xr@f[#1]{#1}
\def\refs#1{\count255=1[\r@fs #1{\hbox{}}]}
\def\r@fs#1{\ifx\und@fined#1\message{reflabel \string#1 is undefined.}%
\nref#1{need to supply reference \string#1.}\fi%
\vphantom{\hphantom{#1}}\edef\next{#1}\ifx\next\em@rk\def\next{}%
\else\ifx\next#1\ifodd\count255\relax\xref#1\count255=0\fi%
\else#1\count255=1\fi\let\next=\r@fs\fi\next}
%

%
\newwrite\ffile\global\newcount\figno \global\figno=1
\def\fig{fig.~\the\figno\nfig}
\def\nfig#1{\xdef#1{fig.~\the\figno}%
\writedef{#1\leftbracket fig.\noexpand~\the\figno}%
\ifnum\figno=1\immediate\openout\ffile=figs.tmp\fi\chardef\wfile=\ffile%
\immediate\write\ffile{\noexpand\medskip\noexpand\item{Fig.\
\the\figno. }
\reflabeL{#1\hskip.55in}\pctsign}\global\advance\figno
by1\findarg}
\def\xfig{\expandafter\xf@g}\def\xf@g fig.\penalty\@M\ {}
\def\figs#1{figs.~\f@gs #1{\hbox{}}}
\def\f@gs#1{\edef\next{#1}\ifx\next\em@rk\def\next{}\else
\ifx\next#1\xfig #1\else#1\fi\let\next=\f@gs\fi\next}
\newwrite\lfile
{\escapechar-1\xdef\pctsign{\string\%}\xdef\leftbracket{\string\{}
\xdef\rightbracket{\string\}}\xdef\numbersign{\string\#}}

\def\writestop{\def\writestoppt{\immediate\write\lfile{\string\pageno%
\the\pageno\string\startrefs\leftbracket\the\refno\rightbracket%
\string\def\string\secsym\leftbracket\secsym\rightbracket%
\string\secno\the\secno\string\meqno\the\meqno}\immediate\closeout\lfile}}
\def\writestoppt{}\def\writedef#1{}
\def\seclab#1{\xdef #1{\the\secno}\writedef{#1\leftbracket#1}\wrlabeL{#1=#1}}
\def\subseclab#1{\xdef #1{\secsym\the\subsecno}%
\writedef{#1\leftbracket#1}\wrlabeL{#1=#1}}
\newwrite\tfile \def\writetoca#1{}
\def\leaderfill{\leaders\hbox to 1em{\hss.\hss}\hfill}
\def\writetoc{\immediate\openout\tfile=toc.tmp
     \def\writetoca##1{{\edef\next{\write\tfile{\noindent ##1
     \string\leaderfill {\noexpand\number\pageno} \par}}\next}}}

%
\catcode`\@=12 
%
\edef\tfontsize{\ifx\answ\bigans scaled\magstep3\else scaled\magstep4\fi} \font\titlerm=cmr10 \tfontsize
\font\titlerms=cmr7 \tfontsize \font\titlermss=cmr5 \tfontsize
\font\titlei=cmmi10 \tfontsize \font\titleis=cmmi7 \tfontsize
\font\titleiss=cmmi5 \tfontsize \font\titlesy=cmsy10 \tfontsize
\font\titlesys=cmsy7 \tfontsize \font\titlesyss=cmsy5 \tfontsize
\font\titleit=cmti10 \tfontsize \skewchar\titlei='177
\skewchar\titleis='177 \skewchar\titleiss='177
\skewchar\titlesy='60 \skewchar\titlesys='60
\skewchar\titlesyss='60
\def\titlefont{\def\rm{\fam0\titlerm}
\textfont0=\titlerm \scriptfont0=\titlerms
\scriptscriptfont0=\titlermss \textfont1=\titlei
\scriptfont1=\titleis \scriptscriptfont1=\titleiss
\textfont2=\titlesy \scriptfont2=\titlesys
\scriptscriptfont2=\titlesyss \textfont\itfam=\titleit
\def\it{\fam\itfam\titleit}\rm}
 \ifx\answ\bigans\else scaled\magstep1\fi
\ifx\answ\bigans\def\abstractfont{\tenpoint}\else
\font\abssl=cmsl10 scaled \magstep1 \font\absrm=cmr10
scaled\magstep1 \font\absrms=cmr7 scaled\magstep1
\font\absrmss=cmr5 scaled\magstep1 \font\absi=cmmi10
scaled\magstep1 \font\absis=cmmi7 scaled\magstep1
\font\absiss=cmmi5 scaled\magstep1 \font\abssy=cmsy10
scaled\magstep1 \font\abssys=cmsy7 scaled\magstep1
\font\abssyss=cmsy5 scaled\magstep1 \font\absbf=cmbx10
scaled\magstep1 \skewchar\absi='177 \skewchar\absis='177
\skewchar\absiss='177 \skewchar\abssy='60 \skewchar\abssys='60
\skewchar\abssyss='60
\def\abstractfont{\def\rm{\fam0\absrm}
\textfont0=\absrm \scriptfont0=\absrms \scriptscriptfont0=\absrmss
\textfont1=\absi \scriptfont1=\absis \scriptscriptfont1=\absiss
\textfont2=\abssy \scriptfont2=\abssys \scriptscriptfont2=\abssyss
\textfont\itfam=\bigit \def\it{\fam\itfam\bigit}\def\footnotefont{\tenpoint}%
\textfont\slfam=\abssl \def\sl{\fam\slfam\abssl}%
\textfont\bffam=\absbf \def\bf{\fam\bffam\absbf}\rm}\fi
\def\tenpoint{\def\rm{\fam0\tenrm}
\textfont0=\tenrm \scriptfont0=\sevenrm \scriptscriptfont0=\fiverm
\textfont1=\teni  \scriptfont1=\seveni  \scriptscriptfont1=\fivei
\textfont2=\tensy \scriptfont2=\sevensy \scriptscriptfont2=\fivesy
\textfont\itfam=\tenit \def\it{\fam\itfam\tenit}\def\footnotefont{\ninepoint}%
\textfont\bffam=\tenbf
\def\bf{\fam\bffam\tenbf}\def\sl{\fam\slfam\tensl}\rm}
\font\ninerm=cmr9 \font\sixrm=cmr6 \font\ninei=cmmi9
\font\sixi=cmmi6 \font\ninesy=cmsy9 \font\sixsy=cmsy6
\font\ninebf=cmbx9 \font\nineit=cmti9 \font\ninesl=cmsl9
\skewchar\ninei='177 \skewchar\sixi='177 \skewchar\ninesy='60
\skewchar\sixsy='60
\def\ninepoint{\def\rm{\fam0\ninerm}
\textfont0=\ninerm \scriptfont0=\sixrm \scriptscriptfont0=\fiverm
\textfont1=\ninei \scriptfont1=\sixi \scriptscriptfont1=\fivei
\textfont2=\ninesy \scriptfont2=\sixsy \scriptscriptfont2=\fivesy
\textfont\itfam=\ninei \def\it{\fam\itfam\nineit}\def\sl{\fam\slfam\ninesl}%
\textfont\bffam=\ninebf \def\bf{\fam\bffam\ninebf}\rm}
%
%

\hyphenation{anom-aly anom-alies coun-ter-term coun-ter-terms}
\def\inv{^{\raise.15ex\hbox{${\scriptscriptstyle -}$}\kern-.05em 1}}

\def\Dsl{\,\raise.15ex\hbox{/}\mkern-13.5mu D} 
\def\dsl{\raise.15ex\hbox{/}\kern-.57em\partial}

 \def\Tr{{\rm Tr}}
\font\bigit=cmti10 scaled \magstep1
\def\lspace{\ifx\answ\bigans{}\else\qquad\fi}
\def\lbspace{\ifx\answ\bigans{}\else\hskip-.2in\fi} 
\def\boxeqn#1{\vcenter{\vbox{\hrule\hbox{\vrule\kern3pt\vbox{\kern3pt
      \hbox{${\displaystyle #1}$}\kern3pt}\kern3pt\vrule}\hrule}}}
\def\mbox#1#2{\vcenter{\hrule \hbox{\vrule height#2in
          \kern#1in \vrule} \hrule}}  
%
 \def\CO{{\cal O}} 
\def\CA{{\cal A}} \def\CC{{\cal C}} \def\CF{{\cal F}} 
\def\CL{{\cal L}} \def\CH{{\cal H}}  
\def\CB{{\cal B}}  \def\CD{{\cal D}} \def\CT{{\cal T}}
\def\e#1{{\rm e}^{^{\textstyle#1}}}

\def\log{{\rm log}}
\def\cos{{\rm cos}}

\def\darr#1{\raise1.5ex\hbox{$\leftrightarrow$}\mkern-16.5mu #1}

\def\half{{\textstyle{1\over2}}} 
\def\roughly#1{\raise.3ex\hbox{$#1$\kern-.75em\lower1ex\hbox{$\sim$}}}

\def\pl#1#2#3{Phys. Lett. {\bf #1B} (#2) #3}

\def\lmp#1#2#3{Lett. Math. Phys. {\bf #1} (#2) #3}
\def\cmp#1#2#3{Comm. Math. Phys. {\bf #1} (#2) #3}

\def\jhep#1#2#3{JHEP {\bf#1}(#2) #3}

\def\IB{\relax\hbox{$\inbar\kern-.3em{\rm B}$}}

\def\ID{\relax\hbox{$\inbar\kern-.3em{\rm D}$}}
\def\IE{\relax\hbox{$\inbar\kern-.3em{\rm E}$}}
\def\IF{\relax\hbox{$\inbar\kern-.3em{\rm F}$}}
\def\IG{\relax\hbox{$\inbar\kern-.3em{\rm G}$}}
\def\IGa{\relax\hbox{${\rm I}\kern-.18em\Gamma$}}
\def\IH{\relax{\rm I\kern-.18em H}}
\def\IK{\relax{\rm I\kern-.18em K}}
\def\IL{\relax{\rm I\kern-.18em L}}
\def\IP{\relax{\rm I\kern-.18em P}}
\def\II{\relax{\rm I\kern-.18em I}}

\def\ndt{{\noindent}}

\def\sssec#1{\ndt$\underline{#1}$}

\def\CA{{\cal A}}
\def\CB{{\cal B}}
\def\CC{{\cal C}}
\def\CD{{\cal D}}

\def\CF{{\cal F}}

\def\CH{{\cal H}}

\def\CL{{\cal L}}
\def\CM{{\cal M}}
\def\CN{{\cal N}}
\def\CO{{\cal O}}

\def\CS{{\cal S}}
\def\CT{{\cal T}}

\def\CX{{\cal X}}

\def\CZ{{\cal Z}}

\def\bC{{\bf C}}

\def\bH{{\bf H}}

\def\bP{{\bf P}}

\def\bR{{\bf R}}
\def\bS{{\bf S}}
\def\bT{{\bf T}}

\def\bV{{\bf V}}

\def\bZ{{\bf Z}}

\def\p{\partial}
\def\pb{\bar{\partial}}



\def\Tr{{\rm Tr}}

\def\sdtimes{\mathbin{\hbox{\hskip2pt\vrule height 4.1pt depth -.3pt
width.25pt\hskip-2pt$\times$}}}

\def\Det{{\rm Det}}


\def\inbar{\,\vrule height1.5ex width.4pt depth0pt}

\def\sdtimes{\mathbin{\hbox{\hskip2pt\vrule height 4.1pt
depth -.3pt width .25pt\hskip-2pt$\times$}}}
\def\a{{\alpha}}

\def\b{{\beta}}
\def\d{{\delta}}
\def\g{{\gamma}}
\def\e{{\epsilon}}
\def\z{{\zeta}}
\def\ve{{\varepsilon}}

\def\m{{\mu}}

\def\u{{\Upsilon}}
\def\l{{\lambda}}
\def\s{{\sigma}}
\def\t{{\sigma}}
\def\vt{{\theta}}

\def\Si{{\Sigma}}

\def\lref{\begingroup\obeylines\lr@f}
\def\lr@f#1#2{\gdef#1{\ref#1{#2}}\endgroup\unskip}

\lref\technique{A.~Schwarz, hep-th/0102182}
\lref\membrane{N. Berkovits,
{\it Toward Covariant Quantization of the Supermembrane},
JHEP 0209 (2002) 051, hep-th/0201151.}
\lref\vanhove{L. Anguelova, P.A. Grassi and P. Vanhove, {\it
Covariant one-loop amplitudes in D=11}, Nucl. Phys. B702 (2004) 269,
hep-th/0408171\semi P.A. Grassi and P. Vanhove, {\it
Topological M Theory from Pure Spinor Formalism}, hep-th/0411167.}
\lref\golos{A.~Gorodentsev and A.~Losev, around late 2000, unpublished.}
\lref\movsch{M.~Movshev and A.~Schwarz, {\it On maximally supersymmetric
Yang-Mills theories}, Nucl. Phys. B681 (2004) 324, hep-th/0311132\semi
M.~Movshev and A.~Schwarz, {\it Algebraic structure of Yang-Mills
theory}, hep-th/0404183.}
\lref\abcd{N.~Nekrasov and S.~Schadchine, {\it ABCD of instantons},
\cmp{252}{2004}{359-391}, hep-th/0404225.}
\lref\FS{L.~D.~Faddeev, S.~L.~Shatashvili, {\it Realization of the Schwinger term in the Gauss law and the possibility of correct quantization of a theory with anomalies}, \pl{167}{1986}{:225-228}}
\lref\pietropeter{P.~A.~Grassi, G.~Policastro , M.~Porrati,  P.~Van~Nieuwenhuizen, {\it Covariant quantization of superstrings without pure spinor constraints}, \jhep{0210}{2002}{054},
hep-th/0112162\semi Y. Aisaka and Y. Kazama, {\it A new first class
algebra, homological perturbation and extension of
pure spinor formalism for superstring}, \jhep{0302}{2003}{017}, hep-th/0212316.}
\lref\opennc{N.~Nekrasov, {\it Lectures on open strings, and noncommutative gauge fields}, hep-th/0203109}
\lref\cdv{
A.~Connes, M.~Dubois-Violette, {\it Yang-Mills algebra}, math.QA/0206205, \lmp{61}{2002}{149-158} }
\def\ihes{{\it Institute des Hautes Etudes Scientifiques, Le
Bois-Marie, Bures-sur-Yvette, F-91440 France}}
\def\itep{{ Institute for Theoretical and
Experimental Physics, 117259 Moscow, Russia}}
\bigskip

\Title{
\vbox{\hbox{IFT-P.061/2005}
\hbox{IHES-P/01/2005}\hbox{ITEP-TH-3/05}\hbox{hep-th/0503075}\hbox{}}}
{\titlefont The Character of Pure Spinors}
\medskip
\vskip 3pt
\centerline{Nathan ~Berkovits$^{1}$ and  Nikita
A.~Nekrasov$^{2}$\footnote{$^{\dagger}$}{On leave of absence from: {\itep}}}
\vskip 1pt
\centerline{$^{1}$ {\it Instituto de F{\'{\i}}sica Te{\'o}rica,
Universidade Estadual
Paulista, Sa{\~o} Paulo, SP 01405, Brasil}}
\centerline{$^{2}$ {\ihes}}
\vskip 12pt
\noindent The character of holomorphic functions
on the space of pure spinors in ten, eleven and twelve dimensions is
calculated. From this character formula, we
derive in a manifestly covariant way various central charges which
appear in the pure spinor formalism for the superstring.
We also derive in a simple way the zero momentum cohomology of the
pure spinor BRST
operator  for the $D=10$ and $D=11$
superparticle.  

\Date{March 2005}
\vfill\eject
\newsec{Introduction}

Five years ago, a new formalism was developed for covariantly
quantizing the superstring \ref\cov{N. Berkovits, {\it
Super-Poincar\'e Covariant Quantization of the
Superstring}, JHEP 04 (2000) 018, hep-th/0001035.}.
This formalism has the new feature that
the worldsheet ghosts are ten-dimensional bosonic spinors $\lambda^\a$
which satisfy the pure spinor constraint $\l^\a \g^m_{\a\b} \l^\b=0$
where $\a$=1 to 16 and $m=0$ to 9. Because of this pure spinor constraint
on the worldsheeet ghost,
the worldsheet antighost $w_\a$
is only defined up to the gauge transformation
$\d w_\a = \Lambda_m (\g^m_{\a\b}\l^\b)$ for arbitrary gauge parameter
$\Lambda_m$. This implies that the worldsheet antighost can only appear in
the combinations
$$N_{mn} = \half \l^\b (\g_{mn})^\a_\b w_\a \quad\quad {\rm and}
\quad\quad J= \l^\a w_\a, $$
which are identified with the Lorentz current and ghost-number current.

To compute scattering amplitudes, one needs to know the OPE's of $N_{mn}$
and $J$ with themselves and with the stress tensor.
Although the single poles in these OPE's are easily determined from
symmetry properties, the central charges need to be explicitly computed.
One method for computing these central charges is to non-covariantly
solve the pure spinor constraint and express $N_{mn}$ and $J$ in terms
of unconstrained variables. Although
this method is not manifestly covariant,
one can easily
verify that the result for the central charges is Lorentz covariant.

Nevertheless, it would be useful to have a manifestly Lorentz-covariant
method for computing these central charges. In addition to the desire
to manifestly preserve as many symmetries as possible,
Lorentz-covariant methods may be necessary if one wants to generalize
the OPE computations in a curved target-space background.

In this paper, it will be shown how to compute the central charges in
a manifestly Lorentz-covariant manner.
By describing the space of pure spinors as the cone over the
$SO(10)/U(5)$
coset and computing the characters of holomorphic functions on this coset,
it will be possible to reinterpret the constrained $\l^\a$ as
an infinite set of unconstrained ghosts-for-ghosts. The central charges
in the OPE's can then be computed by summing the contributions to the
central charges from the infinite set of ghosts-for-ghosts. This
procedure is related to Chesterman's approach to pure spinors \ref
\chesterman{M. Chesterman, {\it Ghost constraints and the covariant
quantization of the superparticle in ten dimensions}, JHEP 0402
(2004) 011, hep-th/0212261.}, and
allows one to complete his construction of a ``second'' BRST
operator. Replacing a constrained pure spinor with 
unconstrained ghosts-for-ghosts has also
been considered in \pietropeter, where a BRST operator was constructed
from only first-class constraints as in the programme of \FS.


In addition to being useful in ten dimensions for covariantly quantizing
the superstring, pure spinors are also useful in eleven dimensions
for covariantly quantizing the $D=11$ superparticle and, perhaps, the
supermembrane \membrane.
Our covariant methods can be easily generalized to compute
characters of functions of pure
spinors in any dimension, in particular eleven and twelve.
Although the Lorentz central charge does not have any obvious meaning
when $D=11$, the
ghost number anomaly does have
a physical meaning and is related to the construction of a
BV action \membrane\ and superparticle scattering amplitudes \vanhove in
$D=11$.

In eleven dimensions, there are two possible definitions of pure spinors
which are either
$$\l \g^I \l =0\quad \quad{\rm or}\quad\quad
\l \g^{IJ} \l =\l \g^{I} \l = 0,$$
where $I,J =0$ to $10$ and ${\l}$ has  $32$ components.
The first definition was used for covariantly quantizing the $D=11$
superparticle and gives rise to a BRST description of linearized $D=11$
supergravity \membrane\ref\cederwall{M. Cederwall, B.E.W. Nilsson
and D. Tsimpis, {\it Spinorial cohomology and maximally
supersymmetric theories}, JHEP 0202 (2002) 009, hep-th/0110069.}.
The second definition was used by Howe in \ref\howe
{P. Howe, {\it Pure Spinors, Function Superspaces and Supergravity
Theories in Ten-Dimensions and Eleven-Dimensions}, Phys. Lett. B273 (1991) 90.}
and can be obtained by dimensionally reducing a $D=12$ pure spinor which
satisifes $\l \g^{MN} \l=0$ for $M,N=0$ to $11$ where the chiral twelve dimensional spinor $\l$ has again
$32$ components.
We explicitly compute the characters using both definitions of $D=11$
pure spinors, however, we have been unable to give a physical interpretation
to characters coming from the second definition.

In section 2 of this paper, non-covariant methods
will be used to solve
the pure spinor constraint in any even dimension and compute
the central charges. In $D$
Euclidean dimensions, these methods manifestly preserve
$U({D}/2)$ invariance. One method requires bosonization and
is a generalization of the parameterization used
in \cov\ for $D=10$ pure spinors. An alternative method uses a
new parameterization of pure spinors which does not
require bosonization.

In section 3,
the character of holomorphic functions on an $SO(10)/U(5)$
coset will be computed using two different covariant methods.
This character formula will then be related to the zero momentum
BRST cohomology
of a $D=10$ superparticle which describes super-Yang-Mills.
Moreover, it will be shown how to generalize
the character formula to describe
superparticle states with non-zero momentum.

In section 4, the character formula will be used to reinterpret the pure
spinor as an infinite set of unconstrained ghosts-for-ghosts,
and the central charges will then be computed by summing the contributions
from the unconstrained ghosts-for-ghosts.

In section 5, we repeat this covariant procedure using a $D=11$ pure
spinor satisfying $\l \g^I \l=0$ where $I=0$ to $10$. The character of
functions on this space of pure spinors will be related to the BRST cohomology
of a $D=11$ superparticle which describes linearized $D=11$
supergravity.
Furthermore, central charges associated with this $D=11$ pure spinor
will be covariantly computed by reinterpreting the pure spinor as
an infinite set of unconstrained ghosts-for-ghosts.

In section 6, we shall compute the character of functions of a $D=11$
pure spinor
satisfying $\l \g^{I}\l=0$ as well as
$\l \g^{IJ}\l=0$, which is equivalent to
computing the character of functions of a $D=12$ pure spinor satisfying
$\l \g^M\l=0$ where $M=0$ to $11$.
In this case, the BRST cohomology associated with the character formula
is complicated and we have not yet found a physical interpretation for it.
We do see, however, an indication for the presence of a self-dual
$D=12$ six-form (field
strength?) in the BRST spectrum, which dimensionally reduces to a
$D=11$ five-form.


\chardef\tempcat=\the\catcode`\@ \catcode`\@=11
\def\cyracc{\def\u##1{\if \i##1\accent"24 i%
    \else \accent"24 ##1\fi }}
\newfam\cyrfam
\font\tencyr=wncyr10
\def\cyr{\fam\cyrfam\tencyr\cyracc}

\newsec{Non-Covariant Computation of Central Charges}

In this section, the central charges in pure spinor OPE's
will be computed by
solving, non-covariantly,
the pure spinor constraint in any even dimension.
The computations are quite simple and will be performed
using two alternative parameterizations of a pure spinor.
The first parameterization requires bosonization of one of
the worldsheet fields, while the second parameterization does not.
We will later rederive these results using
manifestly covariant computations.

In $D=2d$ Euclidean dimensions, a pure spinor $\l^\a$ is constrained
to satisfy $\l^\a (\s^{m_1 ..  m_j})_{\a\b} \lambda^\b =0$ for $0\leq j< d$,
where $m=1$ to $2d$, $\a=1$ to $2^{d-1}$, and $\s^{m_1 ...m_j}_{\a\b}$ is the
antisymmetrized
product of $j$ Pauli matrices. This implies that
$\l^\a \l^\b$ can be written as
\eqn\pures
{\l^\a \l^\b =
{{1}\over{n! ~2^d}}\s_{m_1 ...m_d}^{\a\b} ~(\l^\g \s^{m_1 ... m_d}_{\g\d}
\l^\d)} where
$\l\s^{m_1 ... m_d} \l$ defines an $d$-dimensional complex hyperplane.
This
$d$-dimensional complex hyperplane is preserved up to a phase by a $U(d)$
subgroup of $SO(2d)$
rotations. So projective pure spinors in $D=2d$ Euclidean dimensions
parameterize the
coset space $SO(2d)/U(d)$, which implies that $\l^\a$ has $(d^2 -d +2)/2$
independent complex degrees of freedom \ref\cartan{
E. Cartan, The Theory of Spinors, Dover, New York, 1981\semi P. Budinich and
A. Trautman, The Spinorial Chessboard (Trieste Notes in Physics),
Springer-Verlag, Berlin, 1988.}.

\subsec{Method with bosonization}

One method for computing central charges is to non-covariantly solve
the pure spinor constraint of \pures\ and express $\l^\a$ in terms
of the $(d^2-d+2)/2$ independent degrees of freedom. This was done in \cov\ by
decomposing the $2^{d-1}$ components of $\l^\a$ into $SU(d)\times U(1)$
representations as
\eqn\decone{(\l^{d\over 2} = \g,\quad \l^{{d-4}\over 2}_{[ab]}
= \g u_{[ab]}, \quad
\l^{{{d-8}\over 2}}_{[abcd]} = -{1\over 8} \g u_{[ab} u_{cd]}, \quad
\l^{{{d-12}\over 2}}_{[abcdef]} = -{1\over {48}} \g u_{[ab} u_{cd} u_{ef]},
\quad  ... )}
where the superscript on $\l$ is the $U(1)$ charge,
$\g$ is an $SU(d)$ scalar with $U(1)$ charge $d\over 2$,
and $u_{ab}$ is an $SU(d)$ antisymmetric two-form with $U(1)$ charge $-2$.

Using this decomposition, the naive ghost-number current is $J= \gamma\beta$
and the naive $U(1)$ Lorentz current is
$N_{U(1)} = {d\over 2}\gamma\beta - u_{ab}v^{ab}$
where $\beta$ and $v^{ab}$ are the conjugate momenta for $\gamma$
and $u_{ab}$. However, this naive definition would imply a double pole in
the OPE of $J$ and $N_{U(1)}$ which would violate Lorentz invariance.
As shown in \cov, this double pole can be avoided by bosonizing the
$(\beta,\gamma)$ fields as $(\beta = \p\xi e^{-\phi},\gamma = \eta e^\phi)$
and including terms proportional to $(\p\phi+\eta\xi)$ in
$J$ and $N_{U(1)}$. These terms do not modify the OPE's of
$N$ and $J$ with $\lambda^\alpha$ since $(\p\phi+\eta\xi)$
has no poles with $\l^\a$. However, they do contribute
to the central charges which can be understood as normal-ordering
contributions. The coefficients of these normal-ordering
contributions can be easily determined by requiring that
$N_{mn}$ is a conformal primary field which has no singularities
with $J$.

The formulas for the currents in $D=2d$
are a generalization of the
$D=10$ formulas in \cov\ and are given by
\eqn\firstp{
J = -{5\over 2}\p\phi -{3\over 2} \eta\xi,}
$$
N^{ab} = v^{ab},$$
$$N^b_a =  - u_{ac} v^{bc} + \d_a^b ( {5\over 4} \eta\xi
+ {3\over 4}\p\phi),$$
$$N_{ab} = (d-2) \p u_{ab} + u_{ac} u_{bd} v^{cd} +
u_{ab} ({5\over 2} \eta\xi + {3\over 2} \p\phi),$$
$$T = \half v^{ab} \p u_{ab} - \half \p\phi\p\phi
-\eta\p\xi +\half \p(\eta\xi) + (1-d) \p (\p\phi+\eta\xi),$$
where $T$ is the stress tensor and the worldsheet fields satisfy
the OPE's
\eqn\opeone{\eta(y)\xi(z)\to (y-z)^{-1},\quad
\phi(y)\phi(z) \to -\log(y-z), \quad
v^{ab} u_{cd} \to \d_c^{[a} \d_d^{b]} (y-z)^{-1}.}

\subsec{Method without bosonization}

One can avoid the need for bosonization by using
an alternative
$U(d)$-covariant decomposition of $\l^\a$ in which the pure spinor
is parameterized by an $SU(d)$ vector $e^a$ and an $SU(d)$ antisymmetric
two-form $g^{ab}$. Since a vector and two-form describe
$(d^2 +d)/2$ degrees of freedom and a pure spinor contains only
$(d^2-d+2)$ degrees of freedom, this parameterization will contain
gauge invariances.

In this decompostion, $\l^\a$ splits into the $SU(d)\times U(1)$
representations
\eqn\dectwo{(\l_{{2-d}\over 2}^a = e^a,\quad \l_{{6-d}\over 2}^{[abc]}
= \half e^{[a} g^{bc]}, \quad
\l_{{{10-d}\over 2}}^{[abcde]} = -{1\over 8} e^{[a} g^{bc} g^{de]}, \quad ...)}
where the subscript on $\l$ is the $U(1)$ charge,
$e^a$ is an $SU(d)$ vector with $U(1)$ charge ${2-d}\over 2$,
and $g^{ab}$ is an $SU(d)$ antisymmetric two-form with $U(1)$ charge
$+2$. Note that when $d$ is odd, the $U(d)$ representations in
\decone\ and \dectwo\ are the same (although they are written
in opposite order).
But when $d$ is even, the $U(d)$
representations in \decone\ and \dectwo\ are different since the
pure spinors $\l^\a$ in \decone\ and \dectwo\ have opposite
chirality.

Using the decomposition of \dectwo, the parameterization of $\l^\a$
is invariant under the gauge transformation
\eqn\gaugeone
{\d g^{ab} = \Omega^{[a} e^{b]},}
which in turn has the gauge-for-gauge invariance,
\eqn\gaugetwo{\d \Omega^a = \Lambda e^a.}
Gauge-fixing the invariance parameterized by $\Omega^a$ introduces
a fermionic ghost $\psi^a$ which transforms as an $SU(d)$ vector with
$U(1)$ charge ${d+2}\over 2$ and ghost-number $-1$. (Note that $\l^\a$
has ghost-number one, so $e^a$
has ghost-number one and $g^{ab}$ has ghost-number zero.)
And fixing the
gauge-for-gauge invariance parameterized by $\Lambda$ introduces
a bosonic ghost-for-ghost $r$ which transforms as an $SU(d)$ scalar
with $U(1)$ charge $d$ and ghost-number $-2$.

In terms of these fields and their conjugate momenta,
$$(e^a, f_a;\quad  g^{ab}, h_{ab},\quad  \psi^a, \eta_a; \quad r,s)$$
the worldsheet action is
$$S = \int d^2 z ( f_a \bar\partial e^a +\half h_{ab}\bar\partial g^{ab}
-\eta_a \bar\partial \psi^a + s\bar\partial r),$$
the stress tensor is
$$T=
f_a \partial e^a +\half h_{ab}\partial g^{ab}
-\eta_a \partial \psi^a + s\partial r,$$
the ghost-number current is
\eqn\secondp{J =
e^a f_a
-\psi^a\eta_a -2rs,}
and the Lorentz currents are
$$N_{ab} = h_{ab},$$
$$N^a_b = e^a f_b + g^{ac} h_{bc} + \psi^a \eta_b + \d^a_b (-\half e^c f_c
+\half \psi^c \eta_c + rs),$$
$$N^{ab} = (d-2) \partial g^{ab} -
\half g^{[ab} N^{c]}_c + g^{ab} (g^{cd} h_{cd})
- g^{ad} g^{be} h_{de}.$$
Note that
$N^{ab} = (d-2) \partial
g^{ab}  -\half g^{[ab} \hat N^{c]}_c + g^{ad} g^{be} h_{de}$
where $\hat N^a_b= N^a_b - g^{ac} h_{bc}$.

\subsec{Central charges}

Using either of the two parameterizations of a pure spinor,
the OPE's of the currents in \firstp\ or \secondp\ can
be computed to be
\eqn\OPE{ N_{mn}(y) \l^\a(z) \sim \half {1\over y-z} (\gamma_{mn}\l)^\a, \quad
J(y) \l^\a(z) \sim {1\over y-z} \l^\a,}
$$N^{kl}(y) N^{mn}(z) \sim
{2-d\over  (y-z)^{2}}
(\eta^{n[k} \eta^{l]m}) +
{1\over y-z}(\eta^{m[l} N^{k]n} -
\eta^{n[l} N^{k]m} )
,$$
$$ J(y) J(z) \sim -{4 \over (y-z)^{2}}, \quad J(y) N^{mn}(z) \sim 0, $$
$$N_{mn}(y) T(z) \sim {1\over (y-z)^{2}} N_{mn}(z) ,\quad
J(y) T(z) \sim  {2-2d\over (y-z)^{3}} + {1\over (y-z)^{2}} J(z),$$
$$T(y) T(z) \sim  {\half}{{d(d-1)+2}\over{(y-z)^{4}}} + {2\over{(y-z)^{2}}} T(z) +
{1\over{y-z}} \p T.$$

So the conformal central charge is $d(d-1)+2$, the ghost-number anomaly is
$2-2d$, the Lorentz central charge is $(2-d)$, and the ghost-number central
charge is $-4$.
One can verify the consistency of these charges by considering the
Sugawara construction of the stress tensor
\eqn\sugawara{T = {1\over{2(k+h)}}N_{mn} N^{mn} + {1\over 8} J J +
{{d-1}\over 4}\p J}
where $k$ is the Lorentz central charge, $h$ is the dual Coxeter number
for $SO(2d)$, and the coefficient of $\p J$ has been chosen to give
the ghost-number anomaly $2-2d$.
Setting $k=2-d$, one finds that the $SO(2d)$ current algebra contributes
$(2d-1)(2-d)$ to the conformal central charge and the ghost current
contributes $1+ 3(d-1)^2$ to the conformal central charge. So the
total conformal central charge is $d(d-1)+2$ as expected.

\newsec{The Character of Ten-dimensional Pure Spinors}

In this section, we shall derive a character formula for pure spinors
which will later be used to covariantly compute the central charges.

\subsec{Pure spinors and $Q$ operator}

Consider the space ${\CF}$ of holomorphic functions ${\Psi} ({\l}, {\vt})$,
where
${\l} = ( {\l}^{\a} ) \in {\bS}_{-}$,
$ {\vt} = ( {\vt}^{\a} ) \in {\bS}_{-}$ are the ten dimensional
chiral spinors; $\l$ is bosonic and ${\vt}$ is fermionic. In addition, ${\l}$
obeys the pure spinor constraint ${\l}{\g}^{m} {\l} =0$.
The space of solutions to this constraint, the space of
pure spinors, will be denoted by ${\CM}_{ 10}$. The space ${\CF}$ can
be described as:
\eqn\cf{{\CF} = {\bC} \left[ {\CM}_{10} \right] \otimes {\Lambda}^{\bullet}
{\bS}_{-}}
and is acted upon by the fermionic nilpotent operator
\eqn\qbrst{Q = {\l}^{\a} {{\p}\over{{\p}{\vt}^{\a}}}.}
This operator commutes with the action of the ghost charge operator
\eqn\jzero{K = {\l}{{\p}\over{{\p}{\l}}} +{\vt}{{\p}\over{{\p}{\vt}}}}
and the Lorentz, $Spin(10)$ generators:
\eqn\lorsp{M_{mn} = {1\over 2}{\l}{\g}_{mn} {{\p}\over{{\p}{\l}}} +{1\over 2}
{\vt}{\g}_{mn}
{{\p}\over{{\p}{\vt}}}.}
As a consequence, the ghost charge and the Lorentz group act on the space
${\CH}$ of $Q$-cohomology.
Note that $K$ and $M_{mn}$ differ from the ghost and Lorentz charges of
section 2 since they contain $\theta$ dependence.
The operator $Q$ lowers the fermionic charge $F$,
\eqn\ferch{F = {\vt} {{\p}\over{{\p}{\vt}}} , }by one unit.
The spaces ${\CH}, {\CF}$ are bi-graded by the ghost charge and by the fermionic
charge (the number of ${\vt}$'s):
\eqn\bigra{\eqalign{{\CH} = \bigoplus_{p,q} {\CH}^{p,q} \ & , \qquad {\CF} =
\bigoplus_{p,q} {\CF}^{p,q} \cr
p = K \qquad & \qquad q = 16-F \cr}}
We want to calculate the character of ${\CH}$:
\eqn\chrc{{\chi} ( t, {\s} ) = {\Tr}_{\CH} \left[ \left( -1 \right)^{q} \ t^{K} \ {\bf g} \right]}
where
\def\Si{{\bf L}}
\eqn\lorgen{{\bf g} = \pmatrix{ {\Si}_1 & 0 & 0 & 0 & 0 \cr
0 & {\Si}_2 & 0 & 0 & 0 \cr
0 & 0 & {\Si}_3 & 0 & 0 \cr
0 & 0 & 0 & {\Si}_4 & 0 \cr
0 & 0 & 0 & 0 & {\Si}_5 \cr} = {\exp} \ i {\sigma} \cdot {M}\ , }
where
\eqn\lorgeni{
{\Si}_{a} =  \pmatrix{ {\rm cos}\  {\s}_{a}  & {\sin}\ {\t}_a \cr - {\sin}\ {\s}_a & {\cos}\
{\t}_a \cr}, \qquad a = 1, \ldots 5,}
and ${M} = ( M_{mn} )$, $m,n = 1, \ldots 10$, are the generators of
$Spin(10)$.
This character contains information about the spin content of the $Q$
cohomology. In fact the knowledge of the character
gives the full spin content.

For future use we introduce the notation:
\eqn\ghsch{{\chi}_{p} = \sum_{q = 0}^{16} (-1)^{q} {\Tr}_{{\CH}^{p,q}} (
{\bf g}).}

\subsec{Relating $\chi$ to the character of pure spinors}

We claim that ${\chi}$ can be calculated without knowing the cohomology of
$Q$. The idea is the same as in the calculation of the Euler characteristic
of a manifold. One can do it knowing the Betti numbers, but one can also
do it from using any cell decomposition of the manifold, without actually computing
its cohomology.

Moreover, we claim:
\eqn\ghchsch{{\chi}_{p} = \sum_{q = 0}^{16} (-1)^{q} {\Tr}_{{\CF}^{p,q}} (
{\bf g}).}
This is proven in a standard way:
\eqn\euler{\eqalign{{\Tr}_{{\CF}^{p,q}} =  & {\Tr}_{{\CZ}^{p,q}} + {\Tr}_{{\CF}^{p,q}/
{\CZ}^{p,q}} = \cr
& {\Tr}_{{\CH}^{p,q}} + {\Tr}_{{\CB}^{p,q}} +
{\Tr}_{{\CB}^{p,q+1}}\cr}}
where ${\CZ} = {\rm Ker}Q$, ${\CB} = {\rm Im}Q$, and \euler\ shows that
the contribution of ${\CB}$ drops out from the alternating sum in \ghchsch.

Now the problem of calculating ${\chi}$ is much simpler. In fact,
${\Lambda}^{\bullet}{\bS}_{-}$ contributes
\eqn\thetas{{\Det}_{{\bS}_{-}} \left( 1 - t {\bg} \right)
= \prod_{{\rm even}\ \# \ {\rm of}\  -} \left( 1 - t \ e^{{i\over 2}\left( \pm
{\t}_1 \pm {\t}_2 \pm \ldots \pm {\t}_{5} \right)} \right)}
and it remains to calculate the character of ${\bf g}, t$
acting on the space ${\IH} = {\bC}\left[ {\CM}_{10} \right]$ of
all holomorphic functions on  the space of pure spinors:
$$
Z_{10} ( t, {\t} ) \equiv {\Tr}_{\IH} \left[ t^{K}  e^{ i
{\s}\cdot {M} } \right]
$$
The answer, which will be derived in the following subsections, is:
\eqn\zprt{\mathboxit{Z_{10} ( t,  {\t} ) = {{ \left( 1 -
t^2 V + t^3 S_{+} - t^5 S_{-} + t^6 V - t^8 \right)}\over{ \prod_{{\rm even}
\ {\#} \ {\rm of} \ {-}} \left( 1 - t \ {\exp} {i\over 2} \left( \pm {\t}_1 {\pm} {\t}_2
\ldots {\pm}{\t}_5 \right) \right)}}}}
where (the notations are obvious):
\eqn\chrc{\eqalign{& V = {\Tr}_{{\bV}} \left[ {\bg} \right]  = \sum_{a=1}^{5} 2 \ \cos \  {\t}_a \cr
& S_{+} = {\Tr}_{{\bS}_{+}} \left[ {\bg} \right] = \sum_{{\rm even}
\ {\#} \ {\rm of} \ {+}} {\exp} {i\over 2} \left( \pm {\t}_1 {\pm} {\t}_2
\ldots {\pm}{\t}_5 \right) \cr
& S_{-} = {\Tr}_{{\bS}_{-}} \left[ {\bg} \right]  = \sum_{{\rm even}
\ {\#} \ {\rm of} \ {-}} {\exp} {i\over 2} \left( \pm {\t}_1 {\pm} {\t}_2
\ldots {\pm}{\t}_5 \right) \cr}}
In the following two subsections, we shall first derive the formula of \zprt\
using a reducibility method and will then rederive it using
a fixed-point method. Although the reducibility method is relatively simple
for $D=10$ and is related to the zero-momentum spectrum of
the superparticle,
it becomes more complicated for $D>10$. So
for higher-dimensional
pure spinors, it will be simpler to use the fixed-point method.
We shall then use the formula of \zprt\ to compute various central charges
in a manifestly covariant manner.

\subsec{Reducibility method}

To derive the formula of \zprt\ using the reducibility method, note
that $Z_{10}(t,\t)$ can be written as $P_{10}(t,\t)/Q_{10}(t,\t)$
where $(Q_{10}(t,\t))^{-1}$ is the partition function for an
unconstrained sixteen-component chiral spinor. Furthermore, it will
now be shown that the numerator
\eqn\numer{P(t,\t)=
 1 -
t^2 V + t^3 S_{+} - t^5 S_{-} + t^6 V - t^8 }
 comes from the constraints implied by the pure spinor condition
$\l\g^m\l=0$.

As discussed in the appendix of \ref\particle{N. Berkovits,
{\it Covariant Quantization of the Superparticle using Pure Spinors},
JHEP 0109 (2001) 016, hep-th/0105050.}, the reducibility
conditions for the pure spinor constraint
$\l\g^m\l=0$ can be described
by the nilpotent operator
\eqn\defwq{\widetilde Q = (\wtl\g^m\wtl) b_m
+ c^m (\wtl\g_m f) + (\wtl\g_m\wtl)(j\g^m g) - 2(j_\a \wtl^\a)(g_\b\wtl^\b)}
$$
+ (k\g_m\wtl) r^m + (\wtl\g^m\wtl) s_m t, $$
where $\wtl^\a$ is an unconstrained spinor and
$(c^m,b_m)$, $(g_\a,f^\a)$, $(k^\a,j_\a)$, $(s_m,r^m)$,
and $(u,t)$
are pairs of variables and their conjugate momentum which
have been added to the Hilbert space. In order that $\widetilde Q$ is
fermionic of
ghost-number zero,
the pairs $(c^m,b_m)$, $(k^\a,j_\a)$ and $(u,t)$ are defined to be fermions of
ghost-number $(2,-2)$, $(5,-5)$ and $(8,-8)$ respectively,
and the pairs $(g_\a,f^\a)$ and $(s_m,r^m)$ are defined to
be bosons of ghost-number
$(3,-3)$ and $(6,-6)$ respectively.\foot{In the appendix of \particle,
$\widetilde Q$ was defined to carry ghost-number one, which implied
different ghost numbers for the variables. In the context of this paper,
$\widetilde Q$ is defined to carry ghost-number zero with respect
to the ghost charge $K$ of \jzero. Therefore,
the $t$ dependence in \numer\ counts the ghost number.}
One can check that the coupling of the
variables
$(g_\a,f^\a)$, $(k^\a,j_\a)$, $(s_m,r^m)$,
and $(u,t)$ in \defwq\ describes the reducibility conditions satisfied
by $\wtl\g^m\wtl=0$. For example,
\eqn\couplingf{(\wtl\g^m\wtl)(\g_m\wtl)_\a =0}
implies the coupling to $f^\a$ in \defwq,
\eqn\couplingj{((\wtl\g^n\wtl)\g_n^{\a\b} -2\l^\a\l^\b ) (\g_m\wtl)_\a=0}
implies the coupling to $j_\a$ in \defwq,
\eqn\couplingr{ (\wtl\g_m)_\b
((\wtl\g^n\wtl)\g_n^{\a\b} -2\l^\a\l^\b )=0}
implies the coupling to $r_m$ in \defwq,
and
\eqn\couplingt{ (\wtl\g^m\wtl) (\wtl\g_m)_\b =0}
implies the coupling to $t$ in \defwq.

By comparing \defwq\ with \numer, one can easily check that each term
in $P(t, \t)$ corresponds to a reducibility condition. For example,
the term $-t^2 V$ corresponds to the original pure spinor constraint,
the term $t^3 S_+$ corresponds to the reducibility condition of \couplingf,
the term $-t^5 S_-$ corresponds to the reducibility condition of \couplingj,
etc. So as claimed,
$Z_{10}(t, \t) =P_{10}(t,\t)/Q_{10}(t,\t)$
correctly computes the character formula.

As discussed in \particle, the operator of \defwq\ is related
to the zero momentum spectrum of the $D=10$ superparticle.
It was shown that the super-Yang-Mills ghost, gluon, gluino, antigluino,
antigluon and antighost are described by the states
$(1,c_m, k^\a, g_\a, s_m, u)$ which appear in \defwq.\foot{
The ghost number of these states would be $(0,1,1,2,2,3)$ if one
had defined $\defwq$ to carry ghost-number one as in \particle.}
The fact that these states describe the zero momentum spectrum can
be easily seen by computing the cohomology of \qbrst.
Since the contribution from the partition function for $\vt$ cancels
the denominator of \zprt, the cohomology of \qbrst\ is described by
the numerator $P(t,\vec\t)$ of \numer.

\subsec{The fixed point formula}

In this section we discuss the spaces of pure spinors ${\CM}_{D}$ in various dimensions. 

In the case of even dimension $D = 2d$,
one can use the method of calculating the character which is
well-known in representation theory and is
essentially due to H.~Weyl.
We make use of the fact that the space of pure spinors ${\CM}_{2d}$
is a complex
cone over a compact projective variety ${\CX}_{d} = {\bP}({\CM}_{2d})$. The space
${\IH}_{d}$ of holomorphic functions on ${\CM}_{2d}$ can be then identified
with the space
\eqn\holfn{{\IH}_{d} = \bigoplus_{n=0}^{\infty} {\bH}^{0} \left( {\CX}_{d},
{\CO}(n)\right)}
of holomorphic sections of all powers of the line bundle ${\CL} = {\CO}(1)$, whose
total space is the space ${\CM}_{2d}$ itself, with the blown up origin ${\l}=0$ (the blowup does not
affect the space of holomorphic functions, since the origin has very high codimension). 

The ghost number is precisely $n$, the degree of the bundle.
The character $Z_{D} ( t , { \t})$  can be written as:
\eqn\chrdeg{Z_{D} ( t, {\vec \t} )= \sum_{n=0}^{\infty} t^{n} \sum_i (-1)^i
{\Tr}_{{\bH}^{i} \left( {\CX}_{d},
{\CO}(n)\right)} \left[ {\bg} \right].}
If we are to look at the ghost charge only, then the Riemann-Roch formula
gives immediately:
\eqn\chrdegi{Z_{D} ( t,  0 ) = \int_{{\CX}_{d}}
{1\over{1 - t \ e^{c_{1}({\CL})}}} {\rm Td}_{{\CX}_{d}}}
where ${\rm Td}$ is the Todd genus\foot{It differs from the A-hat genus by the factor
$e^{\half c_1({\CX}_{d})}$. The formula \chrdegi\ can be  derived using
supersymmetric quantum mechanics on ${\CX}_{d}$.}. In terms of the ``eigenvalues"
$x_i$ of the curvature $\cal R$
it is given by:
$$
{\rm Td}_{{\CX}_{d}} = \prod_{i} {x_{i} \over 1 - e^{-x_{i}}} = 1 + {\half}c_{1} ({\CX}_{d})+\ldots
. $$
The full character is given by the equivariant version of the Riemann-Roch
formula.

Now let us utilize the $Spin(D)$ group action on
${\CX}_{d} = {\bP}({\CM}_{2d})$. Consider the space ${\bH}^{0} \left( {\bP}({\CM}_{2d}),
{\CO}(n)\right)$. We can view it as a Hilbert space of a quantum mechanical
problem, where the phase space is ${\CX}$ and the symplectic
form is $n c_{1} ({\CL})$. Then the trace ${\Tr}\left[ {\bf g} \right]$
can be interpreted as the partition function in this quantum mechanical
model with the Hamiltonian ${\t}\cdot M$. It has the path integral
representation, which can be viewed as a infinite-dimensional generalization of
the Duistermaat-Heckmann formula. As such, it is exactly calculable by a
fixed point formula \ref\prseg{A.~Pressley and G.~Segal, Loop
groups, Clarendon Press.} which can be stated in the following generality:
Suppose the space ${\CM}$ is acted on by the group $G$. Suppose also that
$\CM$ is the cone over the base ${\CX}$, ${\CX} = {\CM}/{\bC}^{*}$ and the $G$-action on $\CM$
commutes with scalings in ${\bC}^*$. Fix the generic element ${\bg} \in G$.
Then:
\eqn\fixed{Z ( t , {\bg}) = \sum_{f \in {\CX}^{\bg}}
{1\over{{\Det}_{T_{f}{\CX}}( 1- {\bg} )}} {1\over{1 - t {\chi}_{f}({\bg})}}}
where ${\CX}^{\bg} \subset {\CX}$ is the set of points in $\CX$ which are
invariant under the action of the particular element ${\bg} \in G$.
The fiber ${\CL}_{f}$ of the line bundle ${\CL}$ over $f \in {\CX}$ is the
one-dimensional representation of the subgroup ${\bT} \subset G$, generated by $\bg$: ${\bT} = \overline{\{ {\bg}^n \vert n \in {\bZ}\}}$.
For generic element in a compact Lie group $G$, the subgroup $\bT$ is
isomorphic to the Cartan torus of $G$.
In this case the one-dimensional representation is characterized by the
character ${\chi}_f ({\bg})$ (homomorphism from ${\bT}$ to ${\bC}^*$).

Finally, the tangent space to ${\CX}$ at $f$ is a vector representation of
${\bT}$. The denominator in \fixed\ contains the so-called Weyl denominator,
the determinant of the action of $1 - {\bg}$ on the tangent space to $\CX$
at the fixed point.

In the case $D=11$ the formula \fixed\ cannot be applied directly, since the space ${\CM}_{11}$ is a cone over singular variety. However, by further blowups it can be made smooth and the formula like
\fixed\ can be applied. We discuss this in detail later on. 

\subsec{Even dimensional pure spinors}

In $D = 2d$ dimensions the space ${\bP}({\CM}_{2d})$ of projective pure spinors coincides
with the space of complex structures on the Euclidean space ${\bR}^{2d}$
compatible with the Euclidean structure. This space can be identified
with the coset:
\eqn\projprsp{{\bP}({\CM}_{2d})= SO(2d)/U(d)}
which is a complex manifold of complex dimension
$d(d-1)/2$, over which there is a line bundle $L$,
associated with the $U(1)$ bundle ${\CN}_{2d} = SO(2d)/SU(d) \longrightarrow
{\bP}({\CM}_{2d})$.
The total space of the line bundle $L$ is the moduli space of Calabi-Yau
structures on ${\bR}^{2d}$, i.e. the choices of complex structure and
a holomorphic top degree form, all compatible with the Euclidean structure.
This space is the same as the space of pure spinors in $2d$ dimension.

If the spin zero fields ${\g}$ describe the map
 to ${\CM}_{2d}$ and the spin $(1,0)$ fields $\b$ describe the cotangent
directions to ${\CM}_{2d}$ (the momenta), then the curved
${\b},{\g}$ system
has the Virasoro central charge which
is given by the real dimension of ${\CM}_{2d}$, i.e.:
\eqn\cvir{\mathboxit{c_{\rm Vir} = 2  + d (d - 1)}}

The ghost number anomaly also has a geometric meaning. One can easily show
that the space of projective pure spinors has positive first Chern class of
the tangent bundle:
\eqn\fcc{c_{1} ({\CT}{\bP}({\CM}_{2d})) = (2d-2) c_{1}(L)}
The ghost number anomaly is the first Chern class of the square root of the anticanonical bundle
${\bP}({\CM}_{2d})$. Its topological origin is the same as the shift $k \to k
+ N$ in the level $k$ $SU(N)$ WZW model.

For pure spinors in any dimension, the Virasoro central charge $c_{\rm Vir}$ and
the ghost-number anomaly $a_{\rm ghost}$ are related to
measure factors coming from functional integration \ref\cherkis{
N. Berkovits and
S. Cherkis, {\it Higher-dimensional twistor transforms using pure spinors},
JHEP 0412 (2004) 049, hep-th/0409243.}.
One way to see this is to note that if
$$V=(\lambda)^{n_\l} (\theta)^{n_\theta}$$
is the state of maximal ghost-number
in the pure spinor BRST cohomology,
then $n_\l$ and $n_\theta$ are related to
$c_{\rm Vir}$ and $a_{\rm ghost}$ by
\eqn\topelement{n_\lambda = a_{\rm ghost}
 + \half c_{\rm Vir},\quad n_\theta = N  - \half c_{\rm Vir},}
where $N$ is the number of components of an unconstrained $D$-dimensional
spinor.
For example, when
$D=10$, $n_\l=3$, $n_\theta=5$, $a_{\rm ghost}=-8$, $c_{\rm Vir}=22$,
and $N=16$.
And when $D=11$, $n_\l= 7$, $n_\theta=9$, $a_{\rm ghost}=-16$,
$c_{\rm Vir}=46$, and $N=32$.
The equations of \topelement\ can be understood from functional integration
methods \ref\loops{N. Berkovits, {\it Multiloop amplitudes and vanishing
theorems using the pure spinor formalism for the superstring}, JHEP 0409
(2004) 047, hep-th/0406055.}\vanhove\ since one needs to insert
$\half c_{\rm Vir}$
picture-lowering operators $Y= (C_\alpha \theta^\alpha) \delta
(C_\beta \lambda^\beta)$ to absorb the zero modes of the pure spinor
$\l^\alpha$. So the appropriate non-vanishing inner product for tree amplitudes
is $\langle \l^{n_\l} \theta^{n_\theta} (Y)^{\half c_{\rm Vir}} \rangle$.
Using
the formulas of \topelement, this non-vanishing inner product contains
$N$ $\theta$'s and violates ghost-number by $a_{\rm ghost}$ as expected
from functional integration.

\subsec{Local formula}

The group $Spin(2d)$ acts on ${\bP}({\CM}_{2d})$. Its maximal torus
${\bT}^d$ acts  with $2^{d-1}$ isolated fixed points
${\l}_{{\e}}$, where
\eqn\fxdpt{{\e} = ({\e}_{1}, \ldots, {\e}_{d}) \in {\bZ}_{2}^{d-1}, \qquad {\e}_a = \pm 1 ~~~
{\rm
for}~~~ a
= 1, \ldots d, \qquad \prod_{a} {\e}_a = 1}
The contribution of the fixed point ${\l}_{\e}$ to the character $Z_{2d}(t,
{\t})$, as in \fixed,  is given by:
\eqn\lcltd{\eqalign{Z_{\e} \quad = & \quad {1\over{1 - ts^{-1}_{\e}}} \prod_{1 \leq a < b \leq d}
{1\over{ 1 - e^{i({\e}_{a} {\t}_{a} + {\e}_{b} {\t}_{b})} }}\cr
& \quad s_{\vec\e} = \prod_{a} e^{{i\over 2}{\e}_{a}{\t}_{a}} \cr}}
where the factors
$(1 - e^{i({\e}_{a} {\t}_{a} + {\e}_{b} {\t}_{b})} )^{-1}$
come from the $d(d-1)/2$ $U(d)$ variables $u_{ab}$ which describe
the tangent space near the fixed point.
For the case $d=5$, one can perform the sum over the 16 fixed points
and obtain the formula \zprt\ for $Z_{10}(t,{\s})$.

\ifig\har{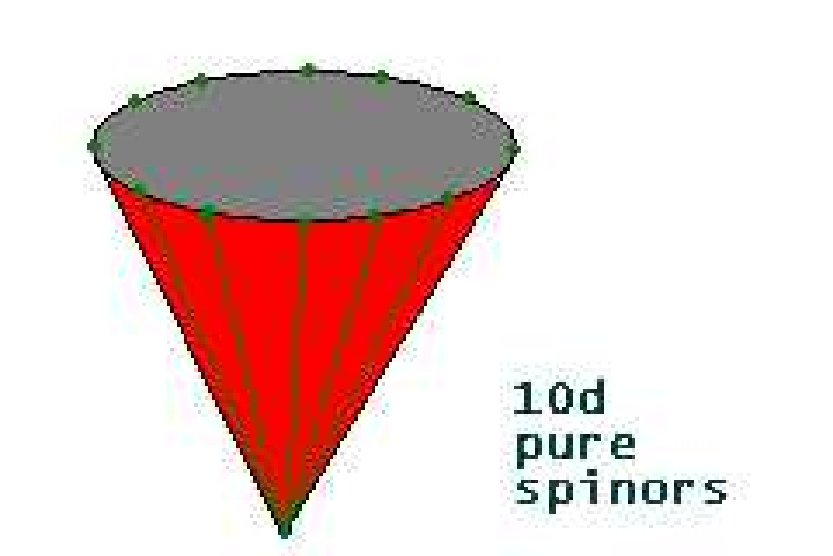}{90}
{The space of pure spinors in ten dimensions and the fixed lines entering \fixed}

\subsec{Non-zero momentum states}

In the study of the superparticle in ten dimensions,
one deals with the $Q$ operator
acting on the space ${\IH}_{X,\theta,\l} $ 
of functions ${\Psi} \left( X, {\theta}, {\l} \right)$, 
where $X$ is the vector of $Spin(10)$, 
${\theta} \in {\bS}_{-}$, is the fermionic spinor, 
and $\l$ is the pure spinor as before:
\eqn\superprt{Q = {\l}\left( {{\p}\over{{\p}{\theta}}} - {\theta}{\gamma}^{m}  {{\p}\over{{\p}{X^{m}}}} \right)}
This operator is Lorentz-invariant. It can be made invariant under the rescaling of $\l$ and $\theta$ provided we scale $X$ with the weight $2$:
\eqn\sscale{{\tilde K} = K + 2 X^{m} {{\p}\over {{\p} X^{m}}} .}

After multiplying the contribution of \thetas\ and \zprt\ from
the $\theta$ and $\l$ variables, one
obtains
\eqn\suprind{{\CZ}_{10} (t, {\t}) = 
{\Tr}_{\IH_{X,\theta,\l}} (-1)^{q} t^{\tilde K} e^{i {\t}\cdot M}
= 
{{P_{10}(t, {\t})}\over{
\prod_{a=1}^{5} ( 1  - t^2 e^{i {\s}_{a}} )(1- t^2 e^{-i {\s}_{a}})}}.}
We see that the degree $2$ term present in
$Z_{10}$ is killed in ${\CZ}_{10}$, 
since the zero momentum state ${\l}{\gamma}{\theta}$ is
$Q$-exact once $X$-dependence is allowed. 
Note that together $\tilde K$ and $M_{2a-1, 2a}$, $a=1, \ldots, 5$ 
form the maximal torus
of the conformal group $SO(1,11)$ in ten (Euclidean) dimensions, 
the symmetry of the classical super-Yang-Mills theory. 
One can arrange the expansion of ${\CZ}_{10}$ in the 
characters of $SO(1,11)$ but we shall
not need it. 
 
\newsec{Covariant Computation of $D=10$ Central Charges}

In this section we shall calculate various central charges of the worldsheet
conformal algebra using the expression \zprt\ for the character.

\subsec{Generators and relations}

Let us forget about the $Spin(10)$ spin content, i.e. let us only
look at the dimensions. In other words, set ${\vec\sigma}=0$.
We get:
\eqn\zprtr{z_{10}(t)= Z_{10}(t,0) = {(1+t)(1+4t+t^2)\over{(1-t)^{11}}}}
Note the relation:
\eqn\invr{z_{10}(t) = (-1)^{11} t^{-8} z_{10}\left( 1/t \right)}
This property of the character is directly related to the ghost number
anomaly. We shall return to it later.

Now suppose we have $N_n$ generators/relations at the ghost number $n$ (this
number
is positive for generators and negative for relations). In other words,
suppose that we have the sequence of free fields:
\eqn\freeflds{{\l}^{\a}, s^{m}, {\s}_{\a}, \ldots}
which represent the sigma model with the target space $\CM$.
These fields are related by:
\eqn\reltns{{s}^{m} = {\l}{\g}^{m}{\l}, \quad {\s}_{\a} =
{s}_{m}\left( {\g}^{m}{\l} \right)_{\a}, \ldots }
and have the ghost numbers $1$, $2$, $3$, $\ldots$. The fields at ghost
number $n$ will have $\vert N_n \vert$ components, and will be bosons for
$N_n > 0$ and fermions for $N_n < 0$.
Then:
\eqn\zprtrr{z_{10} (t)  = \prod_{n=1}^{\infty}  ( 1- t^{n} )^{-N_{n}}}
The multiplicities $N_n$ contain the information about the Virasoro central
charge, as well as the ghost current algebra:
\eqn\cchrgs{\mathboxit{\eqalign{& \half c_{\rm Vir} = \sum_{n} N_{n} \cr
& a_{\rm ghost} = - \sum_{n} n N_{n} \cr
& c_{\rm ghost} = - \sum_{n} n^2 N_{n} \cr}}}
which follows from the consideration of the Lagrangian:
\eqn\lgrns{{\CL} = \int w_{\a} {\pb}{\l}^{\a} - {\pi}_{m}{\pb} s^{m} +
p^{\a} {\pb} {\s}_{\a} + \ldots }
We can easily deduce from \zprtr\zprtrr:
\eqn\rltns{\eqalign{& N_{1} = 16 \cr
& N_{2} = -10 \cr
& N_{3} = 16 \cr
& N_{4} = - 45 \cr
& N_{5} = 144 \cr
& N_{6} = - 456 \cr
& \ldots \cr
& N_{n} \sim (-1)^{n-1}{1\over n} ( 2 + \sqrt{3} )^{n}, \qquad n \to \infty
\cr}}
and also the exact relation:
\eqn\exf{(-1)^{n-1} \left( ( 2 + \sqrt{3} )^{n} + ( 2 - \sqrt{3} )^{n}
\right) = \sum_{d | n } d {\tilde N}_{d}}
where $N_{d} = {\tilde N}_{d}$ for $d > 2$, $N_{1} = 12 + {\tilde N}_1$,
$N_2 = -1 + {\tilde N}_{2}$.

\subsec{Extracting $N_n$}

There is, in fact, an inversion formula for \exf. It utilizes the so-called
M\"obious function, ${\m}(n)$, which is defined as follows:
${\m}(n) = 0$ if $n$ is not integer, or if it is an integer and has one or
more repeated prime factors,
${\m}(1) =1$ and
\eqn\mpr{{\m}(p_1 \ldots p_k ) = (-1)^{k}}
for distinct primes $p_1 , \ldots , p_k$. Then \exf\ implies:
\eqn\exfi{d {\tilde N}_d = \sum_{k} (-1)^{k-1} {\m} \left( {d \over k}
\right) \left( ( 2 + \sqrt{3} )^{k} + ( 2 - \sqrt{3} )^{k}
\right)}
{}From \exfi\ we can get the list of $N_n$'s. Here are the first 30:
\eqn\nnnns{\eqalign{& \{ N_n \} = \cr
\qquad & \{ 16, - 10, 16, - 45, 144, - 456, 1440, -4680,
15600, -52488, \cr
& 177840, -608160, 2095920, -7264080,\cr
\qquad &  25300032, -88517520, 310927680, \cr
&
-1095939000, 3874804560, -13737892896,\cr
& 48829153920, -173949661080, \cr
\qquad &
620963048160, -2220904271040, 7956987570576, -28553733633240, \cr
\qquad &
102617166646800, -369294887482560, \cr
& 1330702217420400, -4800706688284704 \} \cr}}
\subsec{Moments of $N_{n}$'s}

The Mellin transform
\eqn\mobf{\sum_{n=1}^{\infty} {\m}(n) n^{-s} = {1\over{{\z}(s)}}}
of the M\"obius function, which is trivially calculable from the Euler formula
\eqn\eulr{{\z}(s) = \prod_{p} {1\over{1 - p^{-s}}} }
allows to evaluate the regularized moments of the multiplicities $N_n$:
\eqn\mmnts{f(s) = \sum_{n} n^{s+1} N_{n} = 12 - 2^{s+1} + {\tilde f}(s)}
where
\eqn\mmntst{\eqalign{& {\tilde f}(s) = \sum_{n} n^{s+1} {\tilde N}_{n} = \sum_{l,k} (-)^{k-1}{\m}(l ) (k l  )^s
\left(( 2 + \sqrt{3} )^{k} + ( 2 - \sqrt{3} )^{k}
\right) \cr
& \qquad\qquad\qquad = {1\over{{\z}(-s)}} \sum_{k=1}^{\infty} (-)^{k-1} k^s \left( ( 2 + \sqrt{3} )^{k} + ( 2 - \sqrt{3} )^{k}
\right) \cr
& \qquad {\scriptstyle regularization} \cr
& \qquad\qquad\qquad = {{4(2{\pi})^{s} \cos\left({{\pi}s \over 2}\right)}\over{{\z}(s+1)}} \int_{0}^{\infty} {dt \over t}t^{-s} {2 e^t +1 \over e^{2t} + 4 e^t
+1}\cr}}
The last line allows analytic continuation of ${\tilde f}(s)$. The integral
is convergent for $s < 0$. We are interested in the cases $s=-1, 0, 1$.
We have, from \cchrgs:
\eqn\mmnts{\eqalign{& \half c_{\rm Vir} = f(-1) = 11 \cr
& a_{\rm ghost} = - f(0) = - 12 + 2 - {1\over{{\z}(0)}} \left( {{2+\sqrt{3}}
\over{3+ \sqrt{3}}} + {{2-\sqrt{3}}
\over{3 - \sqrt{3}}} \right)  = - 8 \cr
& c_{\rm ghost} = - f(1) = -  12 + 4  - {1\over{{\z}(-1)}} \left( {{2+\sqrt{3}}
\over{(3+ \sqrt{3})^2}} + {{2-\sqrt{3}}
\over{(3 - \sqrt{3})^2}} \right) = - 4\cr}}

\subsec{Ghost number anomaly and central charge without knowing $N_n$}

Consider the general expression of the form we analyzed in the previous section:
\eqn\gncs{\prod_{n} ( 1- t^n )^{-N_n} = {P(t)\over Q(t)}}
where $P$ and $Q$ are some polynomials. We have:
\eqn\lggncs{\sum_{n} N_{n} {\rm log}( 1 - e^{n x} ) = -{\rm log} {P ( e^{x} )
\over Q(e^x ) } }
Since
\eqn\logexp{{\log} ( 1- e^x )  = {\log}(-x) + {x \over 2} +
\sum_{g=1}^{\infty} {B_{2g} \over 2g ( 2g)!} x^{2g}}
where $B_{k}$ are Bernoulli numbers, we have:
\eqn\lggncss{\eqalign{& {\log}(-x) \sum_n N_n + \sum_n {\log}(n) N_n + {x\over 2}
\sum_n n N_n + \cr
& \qquad \qquad \qquad \sum_{g=1}^{\infty}  {B_{2g} \over 2g ( 2g)!} x^{2g} \sum_n
n^{2g} N_{n}  = -{\log}{P(x)\over Q(x)}\cr}}
Expanding the right hand side of \lggncss\ we get all the
even moments of the multiplicity function $N_n$. To get the odd moments besides $\sum_n N_n$ we
ought
to use the more sophisticated machinery of the previous section.

In ten dimensions we have:
\eqn\tend{\eqalign{ & P_{10}(t) = (1+t)(1+4t +t^2), \qquad Q_{10}(t) = (1-t)^{11} \cr
& {\log}P_{10}(e^x)/Q_{10}(e^x) = -11 {\log}(-x) + {\log}(12) -4x - {x^2\over 6} + \ldots \cr}}
and, therefore:
\eqn\ccnth{\eqalign{&  \mathboxit{\half c_{\rm Vir} = \sum_n N_n  = 11} \cr
& \sum_n {\log}(n) N_n = -{\log}(12) \cr
& \mathboxit{ a_{\rm ghost} = - \sum_n n N_n = - 8} \cr
& \mathboxit{c_{\rm ghost}  = - \sum_n n^2 N_n = - 4} \cr}}
(recall that $B_2 = {1\over 6}$).
Even though it is not obviously interesting at the current stage, we also list
the next
few
moments:
\eqn\nxtmmnts{\eqalign{& \sum_n n^4 N_n = 4\cr
& \sum_n n^6 N_n = 4 \cr
& \sum_n n^8 N_n = - {68 \over 3} \cr}}

\subsec{Lorentz current central charge}

Let us go back to the general character \zprt\ and let us specialize:
\eqn\spcz{{\t}_2  = \ldots = {\t}_5 = 0, \qquad e^{i {\sigma}_1 \over 2} = q}
Then
\eqn\chrtsp{\eqalign{& z_{10} ( t, q) \equiv Z_{10}(t, {\t}_{1}, 0,0,0,0) = \cr
&  =  {{(1-t^2)(1 + t (1 + t^2) ( q + q^{-1} ) - 6 t^2
+ t^4)}\over{(1-t q)^7 ( 1 - tq^{-1})^7}} \cr
&\cr
&  \qquad\qquad =  \prod_{n \geq 1, m \in {\bZ}} ( 1- t^{n} q^{m} )^{N_{n,m}} \cr}}
We are interested in
\eqn\chnr{c_{\rm Lorentz} = - \sum_{m,n} {m^2 \over 4} N_{n,m} }
We start with:
\eqn\lggexpns{\sum_{n,m} N_{n,m} {\log} ( 1- t^n q^{m} ) = {\log}\  z_{10}
( t, q)}
from which we derive:
\eqn\lggexpnsi{\eqalign{\sum_{n,m} {m^2 \over 4} N_{n,m} {1\over{( t^{n\over 2} -
t^{-{n \over 2}})^2}} = & - {\p}_{x}^2 \vert_{x=0} {\log}\ z_{10} ( t, e^{x\over 2}) \cr
& - {{2 t ( 2 + 7 t  + 2 t^2)}\over{(1-t)^2(1+4t + t^2)}} = \cr
& - {11 \over 3 ({\log}(t))^2} + {1\over 4} - {131 ({\log}(t))^2 \over 2160} + \ldots
\cr}}
and:
\eqn\mtchexp{\eqalign{& \sum_{n,m} {m^2 \over 4 n^2} N_{n,m} = - {11\over 3}
\cr
& \mathboxit{c_{\rm Lorentz} = - \sum_{n,m} {1\over 4} m^2 N_{n,m} = -3} \cr
& \sum_{n,m} {1\over 4}  m^2 n^2 N_{n,m}  = {13 \over 9} \cr}}
\vfill\eject

\subsec{The ghost number anomaly revisited}

Let us look at the general formula for $Z_{d}$:
\eqn\genfix{Z_{2d}(t, {\vec\t}) = \sum_{{\vec\e}, \prod_{a} {\e}_a = 1 } {1\over{1 - ts^{-1}_{\vec\e}}} \prod_{1 \leq a < b \leq d}
{1\over{ 1 - e^{i({\e}_{a} {\t}_{a} + {\e}_{b} {\t}_{b})} }}}
and let us define the conjugate character by:
\eqn\genfixc{Z_{2d}^{*}(t, {\vec\t}) = \sum_{{\vec\e}, \prod_{a} {\e}_a = -1 } {1\over{1 - ts_{\vec\e}}} \prod_{1 \leq a < b \leq d}
{1\over{ 1 - e^{-i({\e}_{a} {\t}_{a} + {\e}_{b} {\t}_{b})} }}.}
For odd $d$ the conjugate character counts the holomorphic functions on the
space of pure spinors of the opposite chirality. For example, in ten dimensions
this exchanges ${\bS}_{+}$ with ${\bS}_{-}$ in all our formulae above.
For even $d$, $Z_{2d}^{*} =
Z_{2d}$.
Clearly, $Z_{2d}^{*}(t, 0) = Z_{2d}(t, 0)$ for any $d$.

On the other hand, an elementary calculation shows:
\eqn\elemcalc{Z_{2d}^{*}(t, {\t}) = (-1)^{\Delta} \left( t^{-2(d-1)}
Z_{2d}(1/t , {\t}) + \sum_{l=0}^{2(d-2)} t^{l+3-2d} {\tilde
z}_{-l-1}({\t})
\right)}
where ${\Delta} = 1 + d(d-1)/2$ is the complex dimension of ${\CM}_{d}$, and
${\tilde z}_{-l-1}$ is the coefficient at $- t^{-l-1}$ in the
expansion of $Z_{2d}(t, {\vec\t})$ near $t = \infty$. In all our examples
${\tilde z}_{-l-1} = 0$ for $0 \leq l \leq 2d-4$.
On the other hand, the inspection of the Riemann-Roch formula \chrdegi\
gives:
\eqn\rrg{Z_{2d}(t, 0) = (-1)^{\Delta} \left( t^{-{\d}}
Z_{2d}(1/t , 0) + \sum_{l=0}^{2(d-2)} t^{l+3-2d} {\tilde
z}_{-l-1}(0)\right)}
for the manifold ${\CX}$ such that $c_{1}({\CX}) = {\d} c_{1}(L)$. Thus,
${\d} = 2d-2$.

\newsec{Eleven-dimensional Pure Spinors}

In eleven dimensions, $D=11$, there exists a notion of pure spinors which seems
suitable for quantization of superparticle and, perhaps, supermembrane as
well\membrane. The space ${\CM}_{ 11}$ of eleven dimensional pure spinors is the
space of spinors $\l$, obeying
\eqn\eldsp{{\l^\a}{\g_{\a\b}}^{I}{\l^\b} = 0 , \qquad I = 0, \ldots , 10,
\qquad \a=1, \ldots, 32.}
If we decompose ${\l} = ( {\l}_{L}, {\l}_{R} )$ the eleven dimensional
spinor as the sum of the left- and right-chirality ten dimensional spinors,
the equation \eldsp\ reads as:
\eqn\eldspten{\eqalign{& {\l}_L {\g}^{\m} {\l}_{L} +  {\l}_{R}{\g}^{\m}{\l}_{R} = 0 ,
\qquad {\m} = 1, \ldots 10 \cr
& \qquad {\l}_{L}{\l}_{R} = 0 \cr}}
The space of eleven dimensional projective pure spinors ${\bP}({\CM}_{11})$ is the quotient of the space of solutions to \eldspten\ with the point
${\l} = 0$ deleted, by the action of the group ${\bC}^{*}$ of scalings:
${\l} \mapsto s {\l}$, $s \neq 0$. It has (complex) dimension $22$.

\subsec{Global formula}
In this case, the formula for the character (which has the same
arguments as in ten dimensional case, since $Spin(11)$ also has rank five)
is given by:
\eqn\elvdmchr{\eqalign{& \qquad \qquad \qquad \mathboxit{Z_{11}(t, {\t}) \quad =  \quad {{P_{11}(t, { \t})}\over{ Q_{11} (t, {
\t})}}}\cr
& \mathboxit{\eqalign{& \cr & \cr
 P_{11}(t, {\t})\quad  = \quad & \cr
&
\vbox{\hbox{\kern 8pt\vbox{\kern -55pt
\hbox{$\displaystyle \eqalign{
 & \quad  1 -t^2 V + t^4 (V + {\Lambda}^2 V)  \cr &
- t^5 S - t^6 ({\Lambda}^3 V + {\rm Sym}^2 V) + t^7 VS \cr
&  - t^9  VS  + t^{10}
({\Lambda}^3 V + {\rm Sym}^2 V) + t^{11} S  \cr
& -t^{12}(V+{\Lambda}^2 V)  + t^{14} V-t^{16} \cr}$}\kern8pt}\kern8pt}}\cr}} \cr
& \mathboxit{
\quad Q_{11}( t, {\t})  \quad =  \quad {\prod_{\pm\pm\pm\pm\pm}}
\left( 1 - t
e^{{i\over 2}
\left( \pm {\t}_1 \pm {\t}_2 \pm {\t}_3 \pm {\t}_4 \pm {\t}_5 \right)}
\right)\, \quad} \cr}}
where:
\eqn\chrs{\eqalign{& \qquad V = 1+ 2 \sum_{a=1}^5  {\cos}({\t}_{a})
= \sum_{I=0}^{10} e^{y_{I}} \, , \,
 S = \prod_{a=1}^{5} \left( e^{-{i {\t}_{a} \over 2}} +  e^{{i {\t}_{a} \over 2}} \right),  \cr
& {\rm Sym}^2 V = \sum_{I \leq J} e^{y_{I} + y_{J}} \, , \,  {\Lambda}^2 V = \sum_{I < J} e^{y_{I} + y_{J}} \, , \,  {\Lambda}^3 V = \sum_{I < J < K} e^{y_{I} + y_{J} + y_{K}} \cr}}
are the characters of the vector, spinor\foot{Note that $S = \sum_{\pm\pm\pm\pm\pm} e^{{i\over 2}\left( \pm {\t}_{1} \pm {\t}_{2} \pm {\t}_{3} \pm {\t}_{4} \pm {\t}_{5} \right)}$}, symmetric tensor, antisymmetric
rank two tensor, and antisymmetric rank three tensor representations
of $Spin(11)$ respectively.

As in the $D=10$ case discussed earlier, one can derive the above
formula using either the reducibility method or the fixed-point method.
Although the reducibility method is considerably
more complicated than in $D=10$,
its relation to the physical states of the $D=11$ superparticle allows
one to derive the appropriate reducibility conditions.
On the other hand, the fixed-point method is
more straightforward. However, its relation to the physical states
is somewhat indirect.

\subsec{Reducibility method}

To derive the formula of \elvdmchr\ using the reducibility method, note
that $Z_{11}(t,\t)$ can be written as $P_{11}(t,\t)/Q_{11}(t,\t)$
where $(Q_{11}(t,\t))^{-1}$ is the partition function for an
unconstrained 32-component spinor. It will
now be shown that the numerator $P(t,\t)$ of \elvdmchr\
comes from the constraints implied by the $D=11$ pure spinor condition
${\l} {\g}^{I} {\l}=0$ where $I=0$ to 10.

As in the $D=10$ case, the $D=11$ pure spinor constraint
and its reducibility conditions are closely related to the
zero momentum states of the $D=11$ superparticle.
As discussed in the appendix of \membrane,
the reducibility
conditions for the $D=11$ pure spinor constraint
$\l\g^I\l=0$ can be described
by the nilpotent operator
\eqn\complete{\widetilde Q =
\lt\G^I\lt b_{(-1)I}+ c_{(1)}^I(\lt\G_{IJ}\lt u_{(-2)}^J +
\lt\G^J\lt u_{(-2)[IJ]})  }
$$+v_{(2)}^I((\lt\G_I)_\a b^\a_{(-2)} +\lt\G^J\lt b_{(-3)(IJ)})$$
$$
+v_{(2)}^{[IJ]}(\half (\lt\G_{IJ})_\a b^\a_{(-2)}
+\eta_{JK}\lt\G^{KL}\lt b_{(-3)(IL)}
+\lt\G^K\lt b_{(-3)[IJK]})$$
$$+c_{(2)}^\a (-\lt\G^I\lt u_{(-3)I\a}+\half(\lt\G^{IJ})_\a (\lt\G_I)^\b
u_{(-3)J\b}))$$
$$
+\half c_{(3)}^{(JK)}(\lt\G_J)^\a u_{(-3)K\a}+
{1\over 4} c_{(3)}^{[JKL]}(\lt\G_{KL})^\a u_{(-3)J\a}$$
$$+ v_{(3)}^{I\a } b_{(-4)}^{J\b } M_{I\a ~J\b~\g\d}\lt^\g \lt^\d$$
$$+{1\over 4}
u_{(-4)}^{[JKL]}(\lt\G_{KL})^\a c_{(4)J\a} +
\half u_{(-4)}^{(JK)}(\lt\G_J)^\a c_{(4)K\a}$$
$$
+u_{(-5)}^\a (-\lt\G^I\lt c_{(4)I\a}+\half(\lt\G^{IJ})_\a (\lt\G_I)^\b
c_{(4)J\b})$$
$$+b_{(-5)}^{[IJ]}(\half (\lt\G_{IJ})_\a v^\a_{(5)} +
\eta_{JK}\lt\G^{KL}\lt v_{(4)(IL)}
+\lt\G^K\lt v_{(4)[IJK]})$$
$$
+b_{(-5)}^I((\lt\G_I)_\a v^\a_{(5)} +\lt\G^J\lt v_{(4)(IJ)})$$
$$+ u_{(-6)}^I(\lt\G_{IJ}\lt c_{(5)}^J +
\lt\G^J\lt c_{(5)[IJ]}) + b_{(-7)}\lt\G^I\lt v_{(6)I},  $$
where
$ M_{I\a ~J\b~\g\d}$ are fixed coefficients,
and $\lt^\a$ is an unconstrained spinor.

In order that $\widetilde Q$ has ghost-number zero,
the variables
\eqn\onetoa{[1,c_{(1)}^I, v_{(2)}^I, v_{(2)}^{[IJ]}, c_{(2)}^\a, c_{(3)}^{(IJ)},
c_{(3)}^{[IJK]}, v_{(3)}^{I\a},
c_{(4)}^{I\a}, v_{(4)}^{[IJK]},
v_{(4)}^{(IJ)}, v_{(5)}^\a, c_{(5)}^{[IJ]}, c^I_{(5)}, v_{(6)}^I, c_{(7)}],}
carry respectively ghost-number\foot
{In the appendix of \membrane,
$\widetilde Q$ was defined to carry ghost-number one, which implied that
the ghost numbers coincide with the subscript on the variables. In this
paper, $\widetilde Q$ will be defined to carry ghost-number zero with respect
to $K$ of \jzero\ so that the $t$ dependence counts the ghost number.}
\eqn\gntwo{[0,2,4,4,5,6,6,7,9,10,10,11,12,12,14,16].}
As in the $D=10$ case,
one can check that the coupling of the
variables in \complete\
describes the reducibility conditions satisfied
by $\wtl\g^I\wtl=0$.

By comparing \complete\ with \elvdmchr, one can easily check that each term
in $P(t,\t)$ corresponds to a reducibility condition. For example,
the term $-t^2 V$ corresponds to the original pure spinor constraint,
the term $t^4 (V+ \Lambda^2 V)$ corresponds to the reducibility conditions
described by
the second term in \complete, the terms
$- t^5 S -t^6 (\Lambda^3 V + {\rm Sym}^2 V)$ correspond to the
reducibility conditions described by the
second and third lines in \complete, etc.
So, as claimed,
$Z_{11}(t, \t) =P_{11}(t,\t)/Q_{11}(t,\t)$
correctly computes the character formula.

As discussed in \membrane, the operator of \complete\ is related
to the zero momentum spectrum of the $D=11$ superparticle which describes
linearized supergravity.
It was shown that the supergravity ghosts, fields, antifields and antighosts
are described by the states of \onetoa. As in $D=10$,
the fact that these states describe the zero momentum spectrum can
also be seen by computing the cohomology of \qbrst.
Since the contribution from the partition function for $\vt$ cancels
the denominator of \elvdmchr, the cohomology of \qbrst\ is described by
the numerator $P_{11}(t,\t)$ of \elvdmchr.

\subsec{Fixed point method}

The global formula \elvdmchr\ can also
be computed as the sum of $2^5$ local contributions, which
correspond to the fixed points of the action of the element ${\tilde g} =
( {\bf g}, 1)$ on ${\bP}({\CM}_{11})$, where ${\bf g}$ is as in \lorgen.
Unlike the $D=10$ case where the local contribution is straightforward to
calculate, the $D=11$ story is more involved. The difference is the singular
nature of the space of projective pure spinors in $D=11$.
Let us examine the neighbourhood of a fixed point of the element ${\tilde
g}$ action on the space ${\bP}({\CM}_{ 11})$.

First of all, the element ${\tilde g}$ of $Spin(11)$ fixes the point $f$ on
${\bP}({\CM}_{11})$, ${\tilde g}\cdot f = f$, iff it preserves a line $s \cdot {\l}$, $s \neq 0$,
on ${\CM}_{11}$. In other words, there should be a pure spinor ${\l}_{f}$,
which is an eigenvector of $\tilde g$. The fixed points $f \in {\bP}({\CM}_{11})$ are in one-to-one
correspondence with such spinors.

The eigenvectors of $\tilde g$ in the space of all spinors are
in one-to-one correspondence with
the fivetuples of plus and minus signs $\pm\pm\pm\pm\pm$.
The corresponding eigenvalue is
$$
e^{{i\over 2}\left( \pm {\t}_1 \pm {\t}_2 \pm {\t}_3 \pm {\t}_4 \pm {\t}_5
\right)}
$$
We can choose any of these. Thus, we get $2^{5}$ fixed points ${\l}_{\e}$, labeled by
the
vectors ${\e} = ({\e}_1 , {\e}_2 , {\e}_3 , {\e}_4 , {\e}_5) \in {\bZ}_{2}^{5}$, where for each $a=1,\ldots , 5$, ${\e}_a = \pm 1$. The group ${\bZ}_{2}^{5} = W_{B_{5}}/W_{A_{4}}$ is the quotient of the Weyl group of $Spin(11)$ by that of $SU(5)$, the subgroup of $Spin(11)$ preserving the spinor ${\l}_{f}$ . 
Introduce the notations:
\eqn\locw{\eqalign{& w_{a} = e^{i {\e}_a {\t}_a} \cr
& s = \left( \prod_{a=1}^{5} w_{a}\right)^{\half} \cr
& g_{a} = s w_{a}^{-1} \cr
& {\Sigma}_{\pm} = \sum_{a=1}^{5} w_{a}^{\pm 1} \cr}}
Then the contribution of the fixed point ${\e}$ is given by:
\eqn\lclfrml{Z_{11; \e} = {{1 - s^{2} {\Sigma}_{+} + s^4 {\Sigma}_{-}  - s^6}\over{1-t s^{-1}}}
\prod_{m=1}^3
\prod_{a_{1}<a_{2}< \ldots < a_{m}} {1\over{ \left(
1  - w_{a_1} w_{a_2} \ldots w_{a_{m}} \right)}}}
where the various factors appearing in
$Z_{11; \e} $
will be derived below.
Using the formula of \lclfrml, one can derive the formula of
\elvdmchr\ by summing over the 32 fixed points as
$$
Z_{11} (t,\sigma) = \sum_{\e \in {\bZ}_{2}^{5}}
Z_{11; \e} \ .
$$

\sssec{\rm Derivation\ of\ the\ local\ contribution:}

Because of the numerator,
the factor
\eqn\factorel{
(1 - s^{2} {\Sigma}_{+} + s^4 {\Sigma}_{-}  - s^6)
\prod_{m=1}^3
\prod_{a_{1}<a_{2}< \ldots < a_{m}}  {1\over{
1 -e^{i ({\e}_{a_{1}} {\t}_{a_{1}}+ \ldots +{\e}_{a_{m}} {\t}_{a_{m}})}
}}}
in \lclfrml\ differs in form from the typical fixed point
contribution, announced in \fixed. This is because the point ${\l}_{\e}$
is singular, and one needs a more careful analysis of its neighbourhood. We
now present this analysis.

The factors in the denominator in \factorel\
correspond to the parameterization
\eqn\elts{ u_a \sim e^{i{\e}_a{\t}_a}, \quad
u_{ab} \sim e^{i(\e_a\t_a +\e_b\t_b)}, \quad
u_{abc} \sim e^{i(\e_a\t_a +\e_b\t_b +\e_c\t_c)}, }
of the 22-dimensional tangent space near the fixed point.
Since $(u_a, u_{ab}, u_{abc})$ describe 25 parameters, there will be
three constraints on these parameters.
Let us discuss this parameterisation in more detail.
For each split ${\bR}^{11} = {\bR} \oplus {\bC}^{5}$ the $32$-dimensional
space of all spinors can be identified with the space of all $p$-forms,
$p=0,1,2,3,4,5$ on ${\bC}^5$, modulo the factor $\left( {\Lambda}^{5}{\bC}^{5} \right)^{-\half}$:
\eqn\spnrs{{\bS} = \bigoplus_{p=0}^{5} {\Lambda}^{p}{\bC}^{5} \otimes \left( {\Lambda}^{5}{\bC}^{5}
\right)^{-\half}.}
In other words, any spinor can be expressed as
$\sum_{p=0}^5 a_{j_1 ... j_p} \l^{j_1 ... j_p}$ where
$a_{j_1 ... j_p}$ are coefficients and
$\l^{j_1 ... j_p}$ has eigenvalue
$e^{i(\t_{j_1} + ... + \t_{j_p})} (e^{i(\t_1 + ... + \t_5)})^{-\half}$.
It is convenient to choose the eigenvector
${\l}_{\e}$ to be
in the component ${\Lambda}^{0}{\bC}^{5}\otimes \left( {\Lambda}^{5}{\bC}^{5}
\right)^{-\half}$, i.e. $a_{j_1 ... j_p}=0$ for $p>0$.
Now let us consider the infinitesimal variations of this
spinor:
$$
{\l} = {\l}_{f}  \left( 1 + v^{(1)} + v^{(2)} + v^{(3)} + v^{(4)} + v^{(5)} \right) =
{\l}_{f} \left( 1 \oplus v_{a} \oplus v_{ab} \oplus v_{abc} \oplus
v_{abcd} \oplus v_{abcde} \right)
$$
where the $m$-form $v_{a_1 \ldots a_{m}}$ has the eigenvalue
$e^{i ({\e}_{a_{1}} {\t}_{a_{1}}+ \ldots +{\e}_{a_{m}} {\t}_{a_{m}})}$ under
the ${\tilde g}$ action.

The pure spinor constraint \eldsp\ reads in this parameterization,
symbolically:
\eqn\frms{\eqalign{ & v^{(4)} + v^{(2)} v^{(2)} + v^{(1)}
v^{(3)} = 0 \cr & v^{(5)} + v^{(1)} v^{(4)} + v^{(2)}v^{(3)} = 0 \cr & v^{(1)}
v^{(5)} + v^{(2)} v^{(4)} + v^{(3)}v^{(3)} = 0 \cr}} We can solve the first
two equations by:
\eqn\apprx{\eqalign{ & v^{(p)}= {\ve} u^{(p)} ~~~ {\rm for}~~~ p = 1,2,3
\cr & v^{(4)}  = - {\ve}^2 \left( u^{(2)} u^{(2)} + u^{(1)} u^{(3)} \right)
\cr & v^{(5)}  = - {\ve}^2 u^{(2)} u^{(3)} + {\ve}^{3} u^{(1)} \left(
u^{(2)} u^{(2)} + u^{(1)} u^{(3)} \right)\cr}} where $\ve$ is any complex
number. Now the third equation in \frms\ reads (again, symbolically):
\eqn\thrpr{u^{(3)}u^{(3)} + {\ve}\left[ -  u^{(2)} u^{(2)} u^{(2)}  + u^{(1)} u^{(2)} u^{(3)} \right] + {\ve}^{2} \left[ u^{(1)} u^{(1)}
\left(
u^{(2)} u^{(2)} + u^{(1)} u^{(3)} \right) \right] = 0 }
For small ${\ve}$, these equations imply that the infinitesimal deformations
of the pure spinor ${\l}_{\e}$ are parameterized by the arbitrary one-form
$u^{(1)} = ( u_{a})$, the arbitrary two-form $u^{(2)} = ( u_{ab} )$ and
the constrained three-form $u^{(3)} = (u_{abc})$ which obeys
\eqn\thcn{
u^{(3)} u^{(3)} = 0 \Leftrightarrow {\e}^{abcde}u_{abc} u_{def} = 0}
We now have to calculate the character of the ${\tilde g}$ action on the
space of holomorphic functions of $u^{(1)}, u^{(2)}, u^{(3)}$ subject to
\thcn.

The unconstrained scalar
${\l}_f$, the one form $u^{(1)}$ and the two-form $u^{(2)}$ contribute
the factors ${\chi}^{(0)}$, ${\chi}^{(1)}$ and ${\chi}^{(2)}$, respectively:
\eqn\lolcl{\eqalign{& {\chi}^{(0)} = {1\over{1 - t s_{\e}^{-1}}}\cr
& {\chi}^{(1)} = \prod_{a=1}^{5} {1\over{1 - w_{a}}}\cr
& {\chi}^{(2)} = \prod_{1 \leq a < b \leq 5} {1\over{1 - w_{a}w_{b}}}\cr}}
where
\eqn\wwgtht{w_{a} = e^{i {\e}_a {\t}_a}\, .}
Because of the constraint of \thcn, the contribution from $u^{(3)}$ is
more complicated and will be computed below using three alternative methods.

\ifig\onzedim{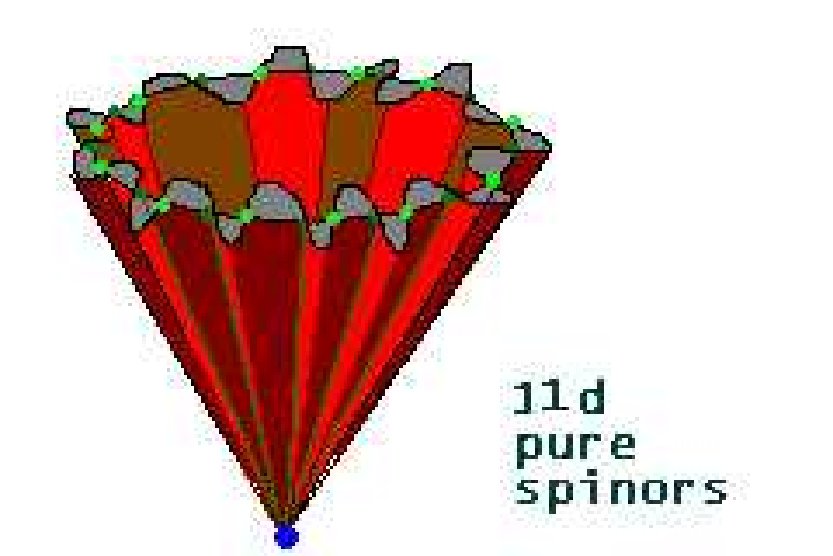}{90}
{The space of pure spinors in eleven dimensions and its singularities}

\sssec{u^{(3)}{\rm  \ contribution\  and \ the \ Grassmanian}\quad Gr(2,5):}

The first method for computing the $u^{(3)}$
contribution uses that \thcn\ is  the
Plucker relation, describing the Grassmanian of two-planes  in
${\bC}^5$ using the decomposable two-forms. In other words, the equation
\thcn\ can be solved by:
\eqn\grs{u_{abc} = {\e}_{abcde} K^{d}_{1} K^{e}_{2}}
where the $2 \times 5$ matrix $K = ( K_{\a}^{d} )$ is defined up to the multiplication by the
$SL_{2}({\bC})$ matrix which acts on the lower index. This gives, by the way, a
simple way to understand the dimension of ${\CM}_{ 11}$. Indeed, the
space of $K$'s is $10$ dimensional. Dividing by $SL_2$ reduces the dimension
by $3$, and together with the dimensions $1$, $5$ and $10$ coming from
${\l}_f$, $u^{(1)}$ and $u^{(2)}$ respectively, we get $23$ announced above.

The contribution ${\chi}^{(3)}$ of $u^{(3)}$ to the character can be therefore expressed as
the $SL_2$ invariant part of the character of the space of functions of the
matrix elements $K_{\a}^{d}$. In other words, we should calculate
\eqn\trgh{
{\Tr}_{{\rm Fun} ( K)} \left( {\tilde g} \times h \right)}
where $h \in SL_{2}({\bC})$, and then average it with respect to $h$. This
is done by integrating over the maximal compact subgroup of $SL_2$, i.e.
$SU(2)$, with respect to the Haar measure on $SU(2)$. Finally, since the
character depends only on the conjugacy class of $h$, the integral reduces
to the integral over the maximal torus of $SU(2)$, i.e. $U(1)$. However, the
measure of integration contains an extra factor, the Weyl-Vandermonde
determinant. For the details of such computations see \abcd.
So, take
$$
h = \pmatrix{ e^{i {\a} \over 2} & 0 \cr 0 & e^{ - {i \a \over 2}} }
$$
The matrix elements $K^{a}_{1}$ are eigenvectors of ${\tilde g} \times h$
with the eigenvalue
$ g_{a} e^{i \a \over 2}$ where
$$ g_{a} = e^{-i{\e}_{a}{\t}_a} s
 $$
while $K^{a}_{2}$ is the eigenvector with the eigenvalue
$
g_{a} e^{-{i \a \over 2}} \ .
$
So \trgh\ is equal to:
\eqn\trghi{{\Tr}_{{\rm Fun} ( K)} \left( {\tilde g} \times h \right) =
\prod_{a=1}^{5} {1\over ( 1 - g_{a} e^{-{i \a \over 2}} ) ( 1 - g_{a} e^{i \a \over
2})}}
The averaging over $SU(2)$ is done by the integral:
\eqn\trghii{{\chi}^{(3)} = {\Tr}_{{\rm Fun} (u^{(3)})} {\tilde g}  =
{1\over{\pi}}\int_{0}^{2\pi}{\rm d}{\a}\ {\rm sin}^{2}\left( {{\a}\over 2}
\right)\ {\Tr}_{{\rm Fun}(K)} {\tilde g} \times h }
which can be evaluated by deforming the contour of integration:
\eqn\trghiii{{\chi}^{(3)}  =
- {1\over{4\pi i}}\oint {{\rm d}z \over z} \left( z - z^{-1} \right)^2 \prod_{a=1}^{5} {1\over ( 1 - g_{a} z ) ( 1 - g_{a} z^{-1})}} and taking the
sum over five residues, at $z = g_{a}$, $a= 1, \ldots ,5$, to get:
\eqn\trghiv{{\chi}^{(3)}  =
{{1 - s^{2} {\Sigma}_{+} + s^4 {\Sigma}_{-}  - s^6}\over{\prod_{a <
b < c} ( 1 - w_a w_b w_c )}}.}
Multiplying the contributions of \lolcl\ and
\trghiv, one obtains the desired formula
of \lclfrml.

\sssec{u^{(3)}{\rm  \ contribution\  from \ an \ ultra-local\ formula}:}

A second method for computing $\chi^{(3)}$ gives rise to an {\it ultra-local}
formula for $\chi^{(3)}$. Although
the five residues of \trghiii\ do not correspond to any
fixed points, the trace ${\Tr}_{{\rm Fun}(u^{(3)})}({\tilde
g})$ has a fixed point expression since
$Gr(2,5)$ is a smooth manifold on which ${\tilde g}$ acts with isolated
fixed points. On the space of $u^{(3)}$'s this action lifts to the action
with fixed projective lines. There are $10 = \pmatrix{5\cr 2} = \pmatrix{5\cr 3}$ of them.
Namely, to each pair $[ab]$, $1 \leq a < b \leq 5$ there corresponds a
two-plane in ${\bC}^{5}$ spanned by the basis vectors ${\be}_a$ and ${\be}_b$.
Dualizing, the fixed points are in one-to-one correspondence with
triples $[abc]$, $1 \leq a < b < c \leq 5$, which label the three-planes
(in the dual ${\bC}^5$) spanned by the coordinate axes ${\be}^a$, ${\be}^b$
and
${\be}^c$. The solution to \thcn, corresponding to $[abc]$ has
$u_{abc} \neq 0$ and the rest of the components of $u^{(3)}$ vanishing.
The contribution of this fixed line to
${\chi}^{(3)}$ is:
\eqn\locloc{{\chi}^{(3)}_{abc} = {1\over{1 - w_{a}w_{b}w_{c}}}
\prod_{d \neq a,b,c}
{1\over{
\left( 1- w_{a}^{-1}w_{d}\right)
\left( 1- w_{b}^{-1}w_{d}\right)
\left( 1- w_{c}^{-1}w_{d}\right)
}}}
The seven factors in the denominator above come from the following
deformations of the $[abc]$ solution to \thcn : the first factor corresponds
to the scaling of the $u_{abc}$ component, while the rest comes from
the homogeneous components of the form: $u_{dbc}/u_{abc}$, $u_{adc}/u_{abc}$ and $u_{abd}/u_{abc}$,
respectively.
Summing over all the triples $[abc]$ we arrive at the expression \trghiv\
for $\chi^{(3)}$:
\eqn\sumr{\sum_{a < b < c} {\chi}^{(3)}_{abc}  = {\chi}^{(3)} , }
as can be verified directly, or using another contour integral
representation\foot{Since $Gr(2,5) = Gr(3,5)$ there is also a
$SL_3$-invariant formalism, where one writes $u_{abc} = M_{a}^{\a}
M_{b}^{\b} M_{c}^{\g} {\ve}_{\a\b\g}$, with $5 \times 3$ matrix $M$ defined
up to the $SL_{3}({\bC})$ action on the index $\a$. The corresponding
projection onto $SU(3)$ invariant states is accomplished by the contour
integral over the two dimensional maximal torus. The residues in this
integral are in one-to-one correspondence with the $[abc]$ triples.}.
Finally, combining all the factors we can write the {\it ultra-local}
formula:
\eqn\loclocf{\eqalign{& Z_{ 11;  abc, {\e}} = \cr
& {1\over{(1 - t s_{\e}^{-1})( 1 - w_{a}w_{b}w_{c})}} \times \cr
& \prod_{f} {1\over{ 1 - w_f}} \times \cr
& \prod_{g < h}{1\over{ 1 - w_g w_h}}\times \cr
&\prod_{d \neq a,b,c}
{1\over{
\left( 1- w_{a}^{-1}w_{d}\right)
\left( 1- w_{b}^{-1}w_{d}\right)
\left( 1- w_{c}^{-1}w_{d}\right)
}}\cr}}
so that:
\eqn\totfx{Z_{11} = \sum_{{\e}, a < b < c} Z_{{11};  abc, {\e}}}
Note that the triples $a<b<c$  together  with $\e$ enumerate precisely the cosets 
$W_{B_{5}}/ W_{A_{1}} \times W_{A_{2}}$, where $W_{A_{1}} \times W_{A_{2}}$ is the Weyl group of $SU(2) \times SU(3)$ -- the stability group of ${\l}_{f} + {\l}_{f}u^{(3)}$.  Also note that the quotient $Spin(11)/SU(2)\times SU(3)$ has  dimension
$44 = 2 \times 22$ which coincides with the dimension of ${\bP}({\CM}_{11})$. In a sense, $Spin(11)/SU(2)\times SU(3)$
is a local model of ${\bP}({\CM}_{11})$ near the fixed lines. 
 
\sssec{ u^{(3)} {\rm \ contribution\ via\ reducibility\ constraints}:}

The third method
for computing the $u^{(3)}$ contribution uses the reducibility properties
of the constraint \thcn\
\eqn\relone{ A_f = \epsilon^{abcde} u_{abc} u_{def} =0  \ . }
Note that $A_f$ satisfies the identity
\eqn\reltwo{  B^e = \epsilon^{abcde} A_a u_{bcd} =0 ,}
and $B^e$ satisfies the identity
\eqn\rethree{ C = B^f A_{f} =0.}
One can easily check that the terms in
the numerator
\eqn\numtwo{1- s^2 \Sigma_{+} + s^4 \Sigma_{-} - s^6}
in \lclfrml\ are related to the reducibility properties of $\epsilon^{abcde}
u_{abc} u_{def}=0$ in
the same manner that the terms in the numerator of \numer\ are related to the
reducibility properties of $\l\g^m\l=0$.
The term $-s^2 \Sigma_+$ comes from the original $A_f$ constraints,
the term $s^4 \Sigma_-$ comes from the $B^e$ constraints, and the term
$-s^6$ comes from the $C$ constraint.

These constraints can also be obtained from the cohomology of
a ``mini''-BRST operator, which acts
on the space of functions of $( u_{abc}, {\vt}_{abc} )$, where $u_{abc}$ are
the components of the constrained bosonic three-form, and ${\vt}_{abc}$ is
an unconstrained fermionic three-form. The operator is:
\eqn\mini{q_{brst} = u_{abc} {{\p}\over{{\p\vt}_{abc}}}}
and squares to zero. The previous statements \relone\reltwo\rethree\
translate to the fact that the cohomology of $q_{brst}$ is spanned
by:
\eqn\chmini{\eqalign{& \qquad\qquad\qquad\qquad\qquad\qquad\qquad 1 \, \cr
& \qquad\qquad\qquad\qquad {\CA}_{f} = {\ve}^{abcde} u_{abc} {\vt}_{def}, \cr
&\qquad\qquad
{\CB}^{e} = {\ve}^{abcde}{\ve}^{a'b'c'd'e'}
{\vt}_{aa'b'}u_{c'd'e'}{\vt}_{bcd}, \cr
& {\CC} = {\ve}^{abca'b'}{\ve}^{defd'e'}{\ve}^{c'f'a''b''c''}u_{abc} u_{def}
{\vt}_{a'b'c'}{\vt}_{d'e'f'}{\vt}_{a''b''c''}.\cr}}
So the numerator of \numtwo\ is related to the cohomology of \mini\
in the same way that the numerator of \numer\ is related to the
cohomology of the $D=10$ superparticle BRST operator.

\subsec{Computation of $D=11$ Central Charges}

To covariantly compute the ghost central charge for a $D=11$ pure spinor,
it is convenient to set $\t=0$ in \elvdmchr\ to obtain the formula
\eqn\elevenzero{z_{11} (t) = Z_{11}( t, 0) = {(1+t)(1+3t+t^2)(1+5t + 10t^2 + 5t^3 +
t^4)\over
(1-t)^{23}} .}
Expansion in ${\rm log}(t)$ as in section 4 yields:
\eqn\ccelevnth{\eqalign{& \quad \mathboxit{\half
c_{\rm Vir} =  \sum_{n} N_n  = 23} \cr
& \mathboxit{ a_{\rm ghost} = - \sum_{n} n N_n = - 16} \cr}}
The ghost and
Lorentz central charges, defined as in the ten dimensional case via the moments of
$N_{n}$ and $N_{n,m}$ turn out to be fractional in eleven dimensions. We don't
have any a priori reasons to expect them to have any meaning, so we shall
not try to explain their failure to be integers.

\newsec{Twelve-dimensional Pure Spinors}
Although the reducibility method for computing the character of
twelve-dimensional
pure spinors is extremely complicated, the fixed-point method is
relatively simple. As will now be shown,
the formula for $Z_{2d+2}(t,\s)$ is related to $Z_{2d}(t,\s)$, which
allows one to obtain the $d=6$ fixed-point formula from the $d=5$ fixed-point
formula of section 3.

\subsec{Recursion relation}

{}From \lcltd\ we can derive a recursion formula which relates $Z_{2d}(t,
{\s} )$ to $Z_{2d+2} (t, {\s}, 0 )$ where
$\s$ has $d$ components. Thus we can read
off the $Spin(2d)$ content for the $2(d+1)$ dimensional pure spinors.
The special feature of the choice ${\s}_{d+1} = 0$ is that the local
contribution \lcltd\ is non-singular, provided the rest of the angles $\s$
is generic. This would not be the case had we set two angles, say
${\t}_{d+1}$ and ${\t}_{d}$ to zero.

For ${\t}_{d+1}=0$ the $2(d+1)$-dimensional character takes the form:
\eqn\hrdmch{Z_{2d+2}(t, {\t} , 0 )  = \sum_{\e}
Z_{2d+2; \e}}
where ${\e}$ is the same as in \fxdpt\ (i.e. has $d$ components),
but now ($s_{\e} = \prod_{a} e^{{i\over 2}{\e}_{a}{\t}_{a}}$):
\eqn\lclhd{\eqalign{Z_{2d+2;  \e} \quad =
& \quad {1\over{1 - ts^{-1}_{\e}}} \prod_{a=1}^{d} {1\over{ 1 - e^{i{\e}_{a} {\t}_{a} }}} \prod_{1 \leq a < b \leq d}
{1\over{ 1 - e^{i({\e}_{a} {\t}_{a} + {\e}_{b} {\t}_{b})} }}\cr
& \quad  \cr}}
\eqn\rcrodd{Z_{2d+2}(t, {\t}, 0) = {1\over{t \prod_{a=1}^{d} \left(
e^{-{i{\t}_a \over 2}} - e^{{i{\t}_a \over 2}} \right) }} \left[ Z_{2d}(t,
{\t} )  - Z_{2d} ( t, - {\t} )\right]}
for odd $d$, and
\eqn\rcreven{Z_{2d+2}(t, {\t}, 0) = {1\over{t \prod_{a=1}^{d} \left(
e^{-{i{\t}_a \over 2}} - e^{{i{\t}_a \over 2}} \right) }} \left[ Z_{2d}(t,
{\t} )  + Z_{2d} ( t, - {\t} ) - 2 \right]}
for even $d$.

\subsec{On to twelve dimensions}

Let us apply the recursion relation \rcrodd\ to the case of twelve
dimensional spinors.
We get, after some simple arithmetics:
\eqn\twlv{Z_{12}(t, {\t} , 0 ) =
{P_{12}(t, {\t} ) \over{Q_{12}(t, {\t})}}}
where
\eqn\numrti{\eqalign{
& \mathboxit{
P_{12}( t, {\t}) \quad = \quad  (1-t^{2})\left( (1+t^{2})^{3} - t^{2} (1 +t^{2}) U + {\half}
t^{3} S \right) {\Gamma} - {\half} t^2 ( 1 + t^2 ) {\Theta} }\cr
& \mathboxit{\qquad \qquad\qquad\qquad\qquad Q_{12} (t, {\t} ) \quad = \quad  \prod_{\e} ( 1 - t s_{\e} ) = {\CD}_{+}{\CD}_{-} \qquad\qquad\qquad}\cr
& \qquad\qquad\qquad\qquad\qquad S = S_{+} + S_{-} \ ,  \cr
& \qquad\qquad\qquad\qquad\qquad  U = 2 \left( 1 + \sum_{a=1}^{5} \ {\cos} {\t}_a \right) \ , \cr
& \qquad\qquad\qquad\qquad\qquad {\Theta} = {\CD}_{+} + {\CD}_{-} \ , \cr
& \cr
& \qquad\qquad\qquad\qquad\qquad {\Gamma} = - {1\over t} {{{\CD}_{+} - {\CD}_{-}}  \over {
S_{+} - S_{-} }} \cr
& \cr
&  \qquad\qquad\qquad\qquad\qquad {\CD}_{\pm} = \prod_{{\e}, \ {\e}_{1} {\e}_{2} {\e}_{3} {\e}_{4} {\e}_{5} = \pm 1} \left( 1 - t s_{\e} \right)\cr }}
For  ${\vec\sigma} = 0$
the character simplifies:
\eqn\numr{
z_{12} (t) = Z_{12}(t,0) =  { 1+ 16 t + 70 t^2 + 112 t^3 + 70 t^4 + 16 t^5 + t^6  \over
(1-t)^{16}}}
If all but one angle vanish, ${\sigma}_{a} = 0 $, $a=2, \ldots , 6$, then the
character assumes the form:
\eqn\numrii{\eqalign{& Z_{12} (t, {\t}_1 , 0,0,0,0,0 ) = {{\rho}_{12}(t,X) \over{ 
\left( 1 - t X + t^2
\right)^{11}}}\cr
& \qquad\qquad\qquad\cr
& \quad {\rho}_{12}(t, X) = 1 - 31 t^2 + 187 t^4 - 330 t^6 + 187 t^8 - 31 t^{10} +
    t^{12} + \cr
    & \quad\quad (5  - 48 t^2 + 55 t^4 + 55 t^6 - 48 t^8 +
          5 t^{10}) tX \cr
          & \quad\quad
+ (5  - 11 t^2 - 11 t^6 +
          5 t^{8})t^2 X^2 + (1 + t^6) t^3 X^3\cr}}
where
$X =2 {\rm  cos} \left( {{\s}_1 \over 2} \right)$.
{}From \numr\numrii\ we derive the ghost number anomaly, the ghost number
current
central charge, Virasoro central charge and the Lorentz current central
charge:
\eqn\crhgs{\mathboxit{\eqalign{& c_{\rm Vir} = 32 \cr
& c_{\rm ghost} = -4 \cr
& c_{\rm Lorentz} = - 4 \cr
& a_{\rm ghost} = - 10 \cr}} \ , }
in agreement with the non-covariant calculation.

\subsec{State  of the art in twelve dimensions. }

We now show the full formula for the twelve dimensional case. The non-trivial part is of course the numerator:
\eqn\nmrts{
\eqalign{
P_{12}(t, {\sigma})  \ = \  & A(t, {\sigma}) + B ( t, {\sigma}) \ ;  \cr
& \cr
A (t, {\sigma}) \ = \ & (1 + t^{22}) -( t^2   + t^{20}) \, U_{[0,0,0,0,1,1]} - (t^4   + t^{18} )\, U_{[-1,1,1,1,1,1]} + \cr &
(t^6   + t^{16} )\times \left( U_{[0,1,1,1,1,2]}+U_{[0,0,0,0,1,3]} \right)+\cr
&  (t^8  + t^{14})\times \left( U_{[0,0,1,1,2,2]}-U_{[0,0,0,0,0,4]} \right) -\cr
& (t^{10}  + t^{12} )\times \left( U_{[1,1,1,1,2,2]}+U_{[0,0,0,2,2,2]}+U_{[0,0,1,1,1,3]} \right) \cr}}
\eqn\bcof{\eqalign{
\qquad B ( t, {\sigma}) \ =\  & ( t^3   + t^{19}  ) \, U_{\half [1,1,1,1,1,3]} - ( t^5   +
          t^{17}  ) \, U_{\half [1,1,1,1,1,5]} - \cr
&  (t^{7}  + t^{15}) \times \left( U_{\half [1,1,3,3,3,3]} + U_{\half [-1,1,1,1,3,5]} \right)
          \cr
& ( t^9    + t^{13}) \times \left(U_{\half [3,3,3,3,3,3]} + U_{\half [-1,1,1,1,1,7]}  \right)+
   2 \  t^{11}  \, U_{\half [1,1,1,3,3,5]}  \cr }}
while the denominator is given by the standard product, as in \numrti. We are using the notation $U_{\l}$ for the character of the irreducible representation of $Spin(12)$ with the highest weight ${\l} = [{\l}_1 , {\l}_{2}, {\l}_{3}, {\l}_{4}, {\l}_{5}, {\l}_{6} ]$, where
$| {\l}_{1} | \leq {\l}_{2} \leq {\l}_{3} \leq {\l}_{4} \leq {\l}_{5} \leq {\l}_{6}$ are integers:
\eqn\chrlam{U_{\l} = e^{i \left( {\l}_{1}{\s}_{1} + \ldots + {\l}_{6}{\s}_{6} \right) } + \ldots }
For example, $U_{[0,0,0,0,0,0]} =1$ corresponds to the trivial representation, $U_{[0,0,0,0,0,1]}$ corresponds to the vector $\bf 12$,
$U_{\half [1,1,1,1,1,1]}$ is the chiral spinor $S_{-}$, and $U_{\half [-1,1,1,1,1,1]}$
is the opposite chirality spinor $S_{+}$.
The dictionary can be continued:
\eqn\rar{\eqalign{&
U_{[0,0,0,0,1,1]} =  {\Lambda}^2 U\cr
&
U_{[-1,1,1,1,1,1]} = {\Lambda}^{6,+}U \cr
&
U_{\half [1,1,1,1,1,3]} =  S_{+} U  -S_{-}  \cr
&
U_{\half [1,1,1,1,1,5]} =S_{-} {\rm Sym}^{2} U - S_{+} U  \cr
&
U_{\half [1,1,3,3,3,3]} + U_{\half [-1,1,1,1,3,5]} =   S_{-} {\Lambda}^{4}U + S_{+}U_{[0,0,0,0,1,2]}-\ldots \cr
&
U_{\half [3,3,3,3,3,3]} + U_{\half [-1,1,1,1,1,7]} =  S_{-} {\Lambda}^{6,+}U + S_{+} {\rm Sym}^3 U -\ldots \cr
&
U_{\half [1,1,1,3,3,5]} = S_{-} U_{[0,0,0,1,1,2]} -\ldots \cr}}
Weyl formula for the character gives:
\eqn\wdsxchr{U_{\l} = {1\over{{\Delta}_{D_{6}}}}
\sum_{{\e} \in {\bZ}_{2}^{5}} \sum_{w \in {\CS}_{6}} (-)^{w} e^{i w ({\l}+{\rho})\cdot {\s}_{\e}}}
where
\eqn\dtlchr{\eqalign{& {\l}+{\rho} = [{\l}_1, {\l}_{2}+1,{\l}_{3}+2,{\l}_{4}+3,{\l}_{5}+4,{\l}_{6}+5] \cr
& w({\m}) = [{\m}_{w(1)},{\m}_{w(2)},{\m}_{w(3)},{\m}_{w(4)},{\m}_{w(5)},{\m}_{w(6)}]\cr
& {\s}_{\e} = [{\e}_1  {\s}_{1} , {\e}_2 {\s}_{2}, {\e}_3  {\s}_{3} , {\e}_4 {\s}_{4},{\e}_5  {\s}_{5} , {\e}_6 {\s}_{6}]\cr
&{\e} = [{\e}_1 , {\e}_2 , {\e}_3 , {\e}_4 , {\e}_{5}, {\e}_{6} ] \in {\bZ}_{2}^{5} ,  {\e}_{i} = \pm 1 \, , \prod_{i=1}^{6} {\e}_{i} = 1 \cr
& {\Delta}_{D_{6}} = \sum_{{\e} \in {\bZ}_{2}^{5}} \sum_{w \in {\CS}_{6}} (-)^{w} e^{i w ({\rho})\cdot {\s}_{\e}} = \cr
& \qquad\qquad \prod_{1 \leq i < j \leq 6} \left( e^{i \left( {\s}_{i} - {\s}_{j} \right) \over 2} - e^{i \left( {\s}_{j} - {\s}_{i} \right) \over 2} \right)\left( e^{i \left( {\s}_{i} + {\s}_{j} \right) \over 2} - e^{-{i \left( {\s}_{j} + {\s}_{i} \right) \over 2}} \right) \cr}}
and the sum in \wdsxchr\ is over the elements $({\e}, w)$ of the Weyl group
of
$Spin(12)$.
The physical interpretation of the field content \nmrts\bcof\ is unclear to us. Presumably the middle component, corresponding to the ghost number $11$, stands for the pair of a physical field and its antifield.  The other components correspond to ghosts, field strengths etc.

The first non-trivial term in \nmrts\ corresponds to the $Q$ cohomology element
${\l}{\Gamma}_{MN} {\vt}$. The degree $3$ term corresponds to
$\left( {\l}{\Gamma}_{MN} {\vt} \right) \left( {\Gamma}^{N} {\vt} \right)^{\a}
-
\left( {\l}{\vt} \right) \left( {\Gamma}_{M} {\vt} \right)^{\a}
$. The self-dual six-form at the degree $4$
deserves further attention.

\vskip 20pt
\noindent{\bf Acknowledgements:} We thank the organizers of Strings'04, the second
Simons workshop at Stony Brook, and the Workshop on Pure Spinors
Formalism at IHES for providing a nice atmosphere during which
some parts of the research on this project have been done. We acknowledge
conversations with M.~Chesterman, P.~Grassi, A.~Khoroshkin,
A.~Losev\foot{Some of the ideas
relating the richness of the
cohomology of the pure spinor BRST operator and the
reducibility of the quadratric pure spinor constraints were discussed in
\golos.}, Y.~Matsuo,
G.~Moore, M.~Movshev\foot{The relation of our ten dimensional character formula to the super-Yang-Mills superfields can
be viewed as an 
independent check of the  
Koszul duality between the (super)-Yang-Mills 
algebra\opennc\cdv\ and the algebra
of functions on ${\CM}_{10}$ , discussed in  \movsch. Also, M.~Movshev has informed
us that M.~Reed has recently constructed a resolvent of the space
of holomorphic functions on ${\CM}_{10}$, from which one can
easily deduce our formula \zprt. We take this as a compliment.} , A.~Okounkov, A.~Polyakov, S.~Shatashvili
and P.~Vanhove. NN is also grateful to NHETC at Rutgers
University, RIMS at Kyoto University, and the Institute for Advanced Study at
Princeton, for hospitality during various stages of preparation of the
manuscript. Research of NN was partly supported by the {\cyr RFFI} grant 03-02-17554
and
by the grant {\cyr NSh}-1999.2003.2 for scientific schools, also by
RTN under the contract 005104 ForcesUniverse.
NB is grateful to CNPq grant 300256/94-9, Pronex grant
66.2002/1998-9 and FAPESP grant 99/12763-0 for partial financial
support and the Fundacao Instituto de Fisica Teorica for their
hospitality.

\vfill\eject
\appendix{A}{On the character calculation in general dimension $2d$}

The sum over fixed points can be done explicitly for $D \leq 10$. Once the formula is established, it can be checked by comparing the residues at the points $t = s_{{\e}}$.

However, for $D=12$ the sum over the $2^5$ fixed points is very complex, and to bring it to the form \nmrts\bcof\ is rather non-trivial.  In general the number of fixed points is $2^{d-1}$ and the complexity of the problem grows exponentially with $N = 2^{d-1}$.

We now present an algorithm, polynomial in $N$, which gives the $t^{k}$ contribution to the
numerator $P_{2d}$ of the general character formula. The number of operations involved
is of the order of $\pmatrix{N \cr k}$.

The idea is to recognize in the sum over the fixed points a part of the sum over the elements of the Weyl group of $Spin(2d)$, as in the general formula \wdsxchr. Indeed, the $2^{d-1}$ fixed points
are in one-to-one correspondence with the cosets $W_{Spin(2d)}/W_{U(d)}$, and the denominator
in the fixed point formula, apart from the $t$-dependent factor, is a subfactor in the Weyl denominator
${\Delta}_{D_{d}}$. The remaining factor is the Weyl denominator for $U(d)$,
$$
{\Delta}_{A_{d-1}} = \prod_{a < b} ( 1 - e^{i ({\e}_a {\s}_{a} - {\e}_{b} {\s}_{b} )} )
$$
which can be introduced at the expense of extending the sum over the cosets to the sum over all of
the Weyl group of $Spin(2d)$.

These considerations lead to the following prescription. Consider the Cartan subalgebra ${\bf h}_{d}$  of $Spin(2d)$. It is a $d$-dimensional vector 
space. Let ${\bf e}_{a}$,
$a = 1, \ldots , d$, denote the orthonormal basis in ${\bf h}_{d}$.  In our previous notations:
$$
 {\bf e}_{a} = [ 0 0 0 \ldots 1 \ldots 0]
 $$
 ($1$ on the $a$'th spot). The positive roots of $Spin(2d)$ are: ${\bf e}_{a} - {\bf e}_{b}$, ${\bf e}_{a} + {\bf e}_{b}$, $1\leq
a < b \leq d$. 
Let $\rho$ denote the half of the sum of the positive roots of $Spin(2d)$:
$$
\rho = \sum_{a=1}^{d} ( a-1) {\bf e}_{a} \ ,
$$
Let $S_{-}$, as before, denote the character of the chiral spinor representation, the one where
${\vt}_{\a}, {\l}_{\a}$ take values:
$$
S_{-} = \sum_{{\e} \in {\bZ}_{2}^{d-1}} e^{{i\over 2} {\e}\cdot {\s}}
$$
We can expand the weights $\m = \half \e$ in the basis ${\bf e}_{a}$: 
\eqn\spinweights{
{\m} = {\half} \sum_{a} {\e}_{a} {\bf e}_{a} \ .}
Let us now introduce an ordering on the weights $\m$:
$$
{\m}_{1} > {\m}_{2} \quad {\rm iff} \quad {\m}_{1}\cdot {\bf e}_{a} = {\m}_{2} \cdot {\bf e}_{a}, \
 a=1,\ldots \ {\ell}-1 , \qquad  {\m}_{1}\cdot {\bf e}_{\ell} > {\m}_{2} \cdot {\bf e}_{\ell}
 $$
 Thus:
 \eqn\orderonweights{
 {\half} [+1+1\dots +1] > {\half} [-1 +1+1\dots +1] > \ldots {\half} [-1 -1 \dots -1]
}
Let ${\l}_{0} =  {\half} [+1+1\dots +1] $ - the maximal (highest) weight.

Then the $t^{k}$ contribution to $P_{2d}$ 
is given by $(-1)^{k}$ times the sum of the characters of the irreducible highest weight representations
$U_{\l}$, with multiplicity $m_{\l}^{(k)}$, where $\l$ obeys: 
${\l} + {\rho} = w ( {\m}_1 + \ldots + {\m}_{k} + {\rho} )$, for some
${\l}_{0} > {\m}_{k} > \ldots {\m}_{1}$, (${\l}_{0}$ is absent because it is cancelled by the $t$-dependent factor in the denominator of the fixed point formula ), 
$w$ is an element of the Weyl group $W_{D_{d}} = W_{A_{d-1}} \sdtimes {\bZ}_{2}^{d-1}$ which makes $\l = [{\l}_1 , \ldots , {\l}_{d}]$ a dominant weight
(i.e.  $|{\l}_{1}| \leq {\l}_{2} \leq \ldots \leq {\l}_{d}$). Finally, if such a $w$ exists, then it is unique, and the
set ${\m}_{1} , \ldots , {\m}_{k}$ contributes to the multiplicity
$m_{\l}^{(k)}$ the amount  $(-1)^w$, which is the signature of the permutation ${\pi}_{w}$ in $W_{A_{d-1}} \equiv {\CS}_{d}$ ($w$ is
a composition   of the permutation ${\pi}_{w}$ and the flip of an even number of signs), otherwise
it contributes $0$ to $m_{\l}^{(k)} $. This gives a prescription for computing
\eqn\finlnmr{\eqalign{ & 
P_{2d} = \sum_{k=0}^{2^{d-1} -2d+2}
\ (-t)^{k} \ \sum_{\l} m_{\l}^{(k)} U_{\l} \cr
&
m_{\l}^{(k)} = 
\sum_{{\m}_{1} < \ldots < {\m}_{k} < {\l}_{0} , w , {\l}+{\rho} = w ( {\m}_{1}+\ldots + {\m}_{k}  + {\rho})} (-1)^{{\pi}_{w}} \cr}}
The sum in \finlnmr\ goes over the $k$-tuples of weights from \spinweights, except for ${\l}_{0}$, ordered according to \orderonweights, and over the elements
$w$ in the Weyl group of $Spin(2d)$, such that the application of $w$ to ${\m}_{1} + \ldots + {\m}_{k} + {\rho}$ gives ${\l} + {\rho}$.

\footatend\vfill\supereject\immediate\closeout\rfile\writestoppt
\baselineskip=14pt\centerline{{\bf References}}\bigskip{\frenchspacing%
\parindent=20pt\escapechar=` \input refs.tmp\vfill\eject}\nonfrenchspacing
\bye